\documentclass{article}
\usepackage{srcltx}
\usepackage{amsmath}
\usepackage{amssymb}
\usepackage{amsfonts}
\usepackage{latexsym}
\usepackage{textcomp}
\usepackage{appendix}
\usepackage{multirow}
\usepackage{booktabs}
\usepackage{url}
\usepackage{array}
\usepackage{marvosym}
\usepackage{graphics}
\usepackage{graphicx}
\usepackage{subfigure}
\usepackage{epsfig}
\newcommand{\N}{{\mathbb N}}

\title{Libor at crossroads: stochastic switching detection using information theory quantifiers}

\author{Aurelio Fern\'andez Bariviera\\ \scriptsize{Department of Business, Universitat Rovira i Virgili, Av. Universitat 1, 43204 Reus, Spain} \\ \scriptsize{\ttfamily aurelio.fernandez@urv.net}   \and M. Bel\'en Guercio \\  \scriptsize{Instituto de Investigaciones Econ\'omicas y Sociales del Sur, UNS-CONICET.} \\  \scriptsize{12 de Octubre y San Juan, B8000CTX Bah\'{\i}a Blanca, Argentina.} \\ \scriptsize{Universidad Provincial del  Sudoeste (UPSO).} \\ \scriptsize{ Alvarado 328, B8000CJH Bah\'ia Blanca, Argentina} \and Lisana B. Martinez \\   \scriptsize{Instituto de Investigaciones Econ\'omicas y Sociales del Sur, UNS-CONICET.} \\  \scriptsize{12 de Octubre y San Juan, B8000CTX Bah\'{\i}a Blanca, Argentina.} \\ \scriptsize{Universidad Provincial del  Sudoeste (UPSO).} \\ \scriptsize{ Alvarado 328, B8000CJH Bah\'ia Blanca, Argentina} \and Osvaldo A. Rosso \\ \scriptsize{Instituto de F\'{\i}sica, Universidade Federal de Alagoas (UFAL). } \\ 
\scriptsize{BR 104 Norte km 97, 57072-970 Macei\'o, Alagoas, Brazil. }\\ \scriptsize{Instituto Tecnol\'ogico de Buenos Aires (ITBA),} \\ \scriptsize{Av. Eduardo Madero 399, C1106ACD Ciudad Aut\'onoma de Buenos Aires, Argentina.}}

\begin{document}
\maketitle

\begin{abstract}
This paper studies the 28 time series of Libor rates, classified in seven maturities and four currencies), during the last 14 years. The analysis was performed using a novel technique in financial economics: the Complexity-Entropy Causality Plane. This planar representation allows the discrimination of different stochastic and chaotic regimes. Using a temporal analysis based on moving windows, this paper unveals an abnormal movement of Libor time series arround the period of the 2007 financial crisis. This alteration in the stochastic dynamics of Libor is contemporary of what press called ``Libor scandal'', i.e. the manipulation of interest rates carried out by several prime banks. We argue that our methodology is suitable as a market watch mechanism, as it makes visible the temporal redution in informational efficiency of the market. 
\end{abstract}%

	\section{Introduction \label{sec:intro}} 

Interest rates no only reflect the time value of money, but also show the tension in the financial market. From the investors' point of view they provide a basic information for making decisions. From the government's point of view they are key elements for effective monetary policy transmission. Consequently fair market conditions in the money market arise as an important issue in political economy.

Libor stands for London Interbank Offered Rate and was created in 1986 by the British Banking Association (BBA). It is one of the most important economic benchmarks, followed closely by those who make financial decisions. According to BBA definition, Libor is ``...the rate at which an individual Contributor Panel bank could borrow funds, were it to do so by asking for and then accepting inter-bank offers in 
reasonable market size, just prior to 11:00 [a.m.] London time''. In fact, Libor rate does not necessarily reflect the cost or price of actual transactions. It is a daily survey conducted by BBA among 16 prime banks, about their fair perception on their own borrowing costs. 
Every London business day, each bank in the Contributor Panel (selected banks from BBA) makes a blind 
submission such that each banker does not know the quotes of the other bankers. 
A compiler, Thomson Reuters, then averages the second and third quartiles. 
This average is published and represents the Libor rate on a given day. 
In other words, Libor is a trimmed average of the expected borrowing rates of leading banks.  
Libor rates has been published for ten currencies and fifteen maturities. 
As it is defined, Libor is expected to be the best self estimate of leading banks borrowing cost at 
different maturities. It is calculated for several currencies and maturities, and the panel composition is not the same for all currencies.

Until 2008, Libor was an uncontested benchmark. However, this situation changed due to a journal publication. Mollenkamp and Whitehouse \cite{MollenkampWhitehouse} published a disruptive article in the Wall Street suggesting that the Libor rate did not reflect what it was expected, \textit{i.e.}, the cost of funding of prime banks. This, and other publications (e.g. \cite{Reuters2012,FTemail}) triggered investigations conducted by the US Department of Justice,  UK Financial Services Authority, EU European Comission and the Swiss Concurrence Commission. In June 2012 Barclays Bank pleaded guilty and accepted a fine of about \$ 480 millions. Other banks were also fined by improper financial conduct.  For a full review of the Libor case from a regulator' point of view, please see Hou and Skeie \cite{HouSkeie}.
 
Only a few papers deal with this topic in academic journals. Most of them uses basic econometric techniques aiming to detect varying differences between the Libor rate and another rate, supposedly not subject to manipulation. Among these papers we find
Taylor and Williams \cite{TaylorWilliams2009}, who documented the detachment of the Libor rate from other market 
rates such as Overnight Interest Swap (OIS), Effective Federal Fund (EFF),  Certificate of Deposits (CDs), 
Credit Default Swaps (CDS), and Repo rates. Snider and Youle \cite{SniderYoule} studied individual quotes in the Libor bank panel and  found that Libor  quotes in the US were not strongly related to other bank borrowing cost proxies. 
Abrantes-Metz  {\it et al.\/} \cite{AbrantesMetz2011} analyzed the distribution of the Second Digits (SDs) of 
daily Libor rates between 1987 and 2008 and, compared it with uniform and Benford's distributions. 
If we take into account the whole period, the null hypothesis that the empirical distribution follows either 
the uniform or Benford's distribution cannot be rejected. However, if we take into account only the period after the subprime crisis, the null hypothesis is rejected. This result calls into question the ``aseptic" setting of Libor.
Monticini and Thornton \cite{Monticini20133} found evidence of Libor under-reporting after analyzing the spread
between 1-month and 3-month Libor and the rate of Certificate of Deposits using the Bai and Perron 
\cite{BaiPerron1998} test for multiple structural breaks. 

Bariviera  {\it et al.\/} \cite{EPJB2015} unveil strange movements in the stochasticity of the 3-month UK Libor, using the Complexity Entropy Causality Plane (CECP). More recently Bariviera {\it et al.\/} \cite{BarivieraRSTA2015} studied the Libor scandal using the Shannon-Fisher plane, giving a new perspective under the lens of local-global information quantifies. 

Our approach greatly expands \cite{EPJB2015}, studying the behavior of the Libor for seven maturities and four currencies using the Complexity Entropy Causality Plane. This study highlights that Libor manipulation was more extensive as originally thought and was more subtle for some maturities. 

The relevance for studying Libor manipulation is that, as stated in the independent study conducted by HM Treasury \cite{WheatleyReport}, more than \$ 300 trilion valued contracts uses Libor as benchmark. This means that the value of syndicated loans, floating rate notes and interest rate swaps were affected. Even more, many mortgages have their interests linked to Libor evolution. As a consequence borrowers (mostly families) were directly affected by this unfair behavior. 

The rest of the paper is structured as follows. Section \ref{sec:InfoTheory} describes the methodology. Section \ref{sec:Data} details the data under analysis. Section \ref{sec:Results} comments the main findings of our study and, finally Section \ref{sec:Conclusions} concludes.

\section{Information theory quantifiers \label{sec:InfoTheory}}

Many economic data are recorded as a sequence of measurements equally spaced in time. This kind of data, commonly referred as time series, are usually the starting point for economic analysis. When the data are abundant, the number of adequate quantitative techniques increases. In particular, econophysics methods, as the one applied in this article, are innovative and appropriate to shed light on economic phenomena. In many cases, econophysics complement the limitations of traditional econometric techniques.

In this line, information-theory-derived quantifiers can help to extract relevant information from financial time series. The use of information quantifiers in economics is not new, but infrequent. The origins can be traced back to Theil and Leenders \cite{TheilLeenders65}, who use entropy to predict short-term price fluctuations in the Amsterdam Stock Exchange. \cite{Fama65entropy} and \cite{Dryden68} replicate the same technique for the New York Stock Exchange and the London Stock Exchange respectively. \cite{PhilippatosNawrocki73} analyzes the proportion of securities with positive, negative and null returns on the American Stock Exchange using information theory methods and conclude that this proportions are dependent on the previous day and is not significantly influenced by the proportion of untraded securities. \cite{PhilippatosWilson74} proposes the average mutual information or shared entropy as a proxy of systematic risk. This technique was remained unused until recent years. For example, \cite{Risso08} uses entropy and symbolic time series analysis in order to relate informational efficiency and the probability of having an economic crash. Later, \cite{Risso09} uses Shannon entropy to rank the informational efficiency of several stock markets around the world. \cite{OrtizCruz12} uses multiscale entropy analysis to analyze the evolution of the informational efficiency of crude oil prices.

\subsection{Shannon entropy}
When studying dynamical systems, the discrimination of the presence of correlations in time series, emerges as one key task. 
Given a time series, one of the most natural measures of disorder, and thus absence of correlation, is Shannon entropy \cite{book:shannon1949}. Given a discrete probability distribution $P=\{p_i: i=1,\cdots M\}$, Shannon entropy is defined as:
\begin{equation}
\label{Shannon}
S[P]~=~ -\sum_{i=1}^{M} p_i \log(p_i) 
\end{equation}
This formula measures the information embedded into the physical process decribed by $P$. It is a bounded function in the interval $[0, \log(M)]$. $S[P]=0$ means that one of the states $p_{i^*}=1$ and the remaining $p_i=0$ for $i\neq i^*, \forall i \in M$. In other words, null entropy means full certainty about the system's outcome. On the other extreme, if $S[P]= \log(M)$, our knowledge about the system is minimal, meaning that all states are equally probable. 
Even though entropy can describe globally the level of order/disorder of a process, the analysis of time series using solely Shannon entropy could be incomplete \cite{FeldmanCrutchfield98}. The reason is that an entropy measure does not quantify the degree of structure or patterns present in a process. Consequently, a measure of statistical complexity is necessary in order to characterize the system.

\subsection{Statistical complexity}

Although Shannon entropy is a good measure of the order of a physical system, it has limitations. An additional measure in order to measure the hidden structure of the process is needed in order to fully characterize dynamical systems: an statistical complexity measure.
A family of statistical complexity measures, based on the functional form developed by \cite{LMC95}, is defined in \cite{Martin2003,Lamberti2004119} as:
\begin{equation}
{\cal C}_{JS}= {\cal Q}_{J}[P,P_e] \cdot {\cal H} [P]
\label{eq:complexity}
\end{equation}
where ${\cal H} [P]=S[P]/S_{\max}$ is the normalized Shannon entropy, $P$ is the discrete probability distribution of the time series under analysis, $P_e$ is the uniform distribution and ${\cal Q}_{J}[P,P_e]$ is the so-called disequilibrium. This disequilibrium is defined in terms of the Jensen-Shannon divergence, which quantifies the difference between two probability distributions. Mart\'{\i}n, Plastino and Rosso \cite{paper:martin2006} demonstrates the existence of upper and lower bounds for generalized statistical complexity measures such as ${\cal C}_{JS}$ . Additionally, as highlighted in \cite{Soriano2011a}, the permutation complexity is not a trivial function of the permutation entropy because they are based on two probability distributions. A complete discussion about this measures and details about their calculation is in \cite{Zunino2010a}.

\subsection{Bandt-Pompe symbolization method} 

In order to evaluate this quantifiers, a symbolic technique should be selected in order to obtain the appropriate probability distribution function. Following \cite{Zunino2010a,ZuninoCausality10,ZuninoPermutation11,RossoNoise07}, we use the Bandt and Pompe \cite{BandtPompe02} permutation method, because it is the single symbolization technique that considers time causality. This methodology requires only weak stationarity assumptions.

The appropriate symbol sequence  arises naturally from the time series.  ``Partitions'' are devised  by comparing the order of neighboring relative values rather than by apportioning amplitudes according to different levels. No model assumption is needed because Bandt and Pompe method makes partitions of the time series and orders values within each partition. 
Given a time series ${\mathcal S}(t) = \{ x_t ; t = 1, \cdots , N \}$,
an embedding dimension $D > 1, D \in \N$, and an embedding delay $\tau, \tau \in \N$, the BP-pattern of order $D$ generated by
\begin{equation}
\label{eq:vectores}
s \mapsto \left(x_{s-(D-1)\tau},x_{s-(D-2)\tau},\cdots,x_{s-\tau},x_{s}\right) 
\end{equation}
is the one to be considered.
To each time $s$, BP assign a $D$-dimensional vector that results from the evaluation of the
time series at times $s - (D - 1) \tau, s-(D-2)\tau, \cdots , s - \tau, s$.
Clearly, the higher the value of $D$, the more information about ``the past'' is incorporated
into the ensuing vectors.
By the ordinal pattern of order $D$ related to the time $s$, BP mean the permutation
$\pi = (r_0, r_1, \cdots , r_{D-1})$ of $(0, 1, \cdots ,D - 1)$ defined by
\begin{equation}
\label{eq:permuta}
x_{s-r_{D-1} \tau} \le  x_{s-r_{D-2} \tau} \le \cdots \le x_{s-r_{1} \tau}\le x_{s-r_0 \tau}. % \ .
\end{equation}
In this way the vector defined by Eq. (\ref{eq:vectores}) is converted into a definite symbol $\pi$.
So as to get a unique result BP consider that $r_i < r_{i-1}$ if $x_{s-r_{i} \tau} = x_{s-r_{i-1} \tau}$.
This is justified if the values of ${x_t}$ have a continuous distribution so that equal values are
very unusual.

For all the $D!$ possible orderings (permutations) $\pi_i$ when  embedding dimension is $D$, their associated relative
frequencies can be naturally computed according to the number of times this particular order sequence is found in the time series,
divided by the total number of sequences,
\begin{equation}
\label{eq:frequ}
p(\pi_i)= \frac{\sharp \{s|s\leq N-(D-1)\tau ; (s) \quad \texttt{has type}~\pi_i \}}{N-(D-1)\tau} %\ .
\end{equation}
In the last expression the symbol $\sharp$ stands for ``number''.
Thus, an ordinal pattern probability distribution $P = \{ p(\pi_i), i = 1, \cdots, D! \}$ is obtained from the time series.

As we mention previously, the ordinal-pattern's associated PDF is invariant with respect to nonlinear monotonous
transformations. Accordingly, nonlinear drifts or scalings artificially introduced by a
measurement device will not modify the quantifiers' estimation, a nice property if one deals with experimental data (see i.e. \cite{Saco2010}).
These  advantages make the BP approach more convenient than conventional methods based on range partitioning.
Additional advantages of the  method reside in (i) its simplicity (we need  few parameters: the pattern length/embedding
dimension $D$ and the embedding delay $\tau$) and (ii) the extremely fast nature of the pertinent calculation-process \cite{paper:keller2005}.
The BP methodology can be applied not only  to time series representative
of low dimensional dynamical systems but also to any type of time
series (regular, chaotic, noisy, or reality based).
In fact, the existence of an attractor in the $D$-dimensional phase space in not assumed.
The only condition for the applicability of the BP method is  a very
weak stationary assumption (that is, for $k=D$, the probability
for $x_t < x_{t+k}$ should not depend on $t$ \cite{BandtPompe02}).
The selected pattern length should fulfill $N \gg D!$ , in order to obtain reliable quantifiers .

\subsection{The Complexity Entropy Causality Plane}

When the Shannon entropy and the statistical complexity measures defined before are computed using the \cite{BandtPompe02} symbolization technique, the quantifiers are named permutation entropy and permutation statistical complexity. Both quantifiers can be represented in a Cartesian plane, forming the Complexity Entropy Causality Plane (CECP). This planar representation was introduced in efficiency analysis in \cite{Zunino2010a} and was successfully used to rank efficiency in stock markets \cite{ZuninoCausality10}, commodity markets \cite{ZuninoPermutation11}, and to link informational efficiency with sovereign bond ratings \cite{Zunino2012}. Given the power of the CECP for the discrimination of random and chaotic signals, its application goes across disciplines. For example, \cite{SerinaldiZuninoRosso2013} studies the daily stream flow of United States rivers, and \cite{Zanin2012} reviews the main biomedical and econphysical applications of this methodology.

\section{Data \label{sec:Data}}

We analize the Libor rates in British Pounds (GBP), Euro (EUR), Swiss Franc (CHF) and Japanese Yen (JPY), for the following seven maturities: overnight (O/N), one week (1W), one month (1M), two months (2M), three months (3M), six months (6M) and twelve months (12M). The data coverage is from 02/01/2001 until 06/10/2015, for a total of 3851 datapoints. All data were retrieved from Datastream. 

We computed the permutation entropy and permutation statistical complexity for $D=4$, using daily values ($\tau=1$). In order to assess the changes in the dynamical process that generates Libor time series, we used sliding windows. The sliding window approach works as follows: we compute the information quantifiers for the first 300 datapoints, then we move forward 20 datapoints ($\delta=20$) and compute again the quantifiers for the next 300 datapoints. We continue in this way until the end of the data. Using this procedure, we obtained 177 windows, each one spanning slightly more than a year ($\approx 13$ months)

\section{Results \label{sec:Results}}
The results of the permutation entropy and statistical complexity are displayed in cartesian planes called Complexity Entropy Causality Planes. This graphical representation allows the discrimination of stochastic and chaotic dynamics, as described in \cite{RossoNoise07}. According to the classical financial literature, prices in a competitive market should follow a memoryless stochastic process \cite{Samuelson65}. Thus, if Libor is freely set, without exogenous altering forces, it should approximately follow a random walk. In this situation, permutation entropy is maximized and permutation statistical complexity is minimized. We can safely say that, the closer the quantifiers to the point $(1,0)$, the more informational efficient the market is.

\begin{figure}[!ht]
    \centering
{
        \subfigure{%
            \label{fig:CECPgbpON}
            \includegraphics[width=0.45\textwidth]{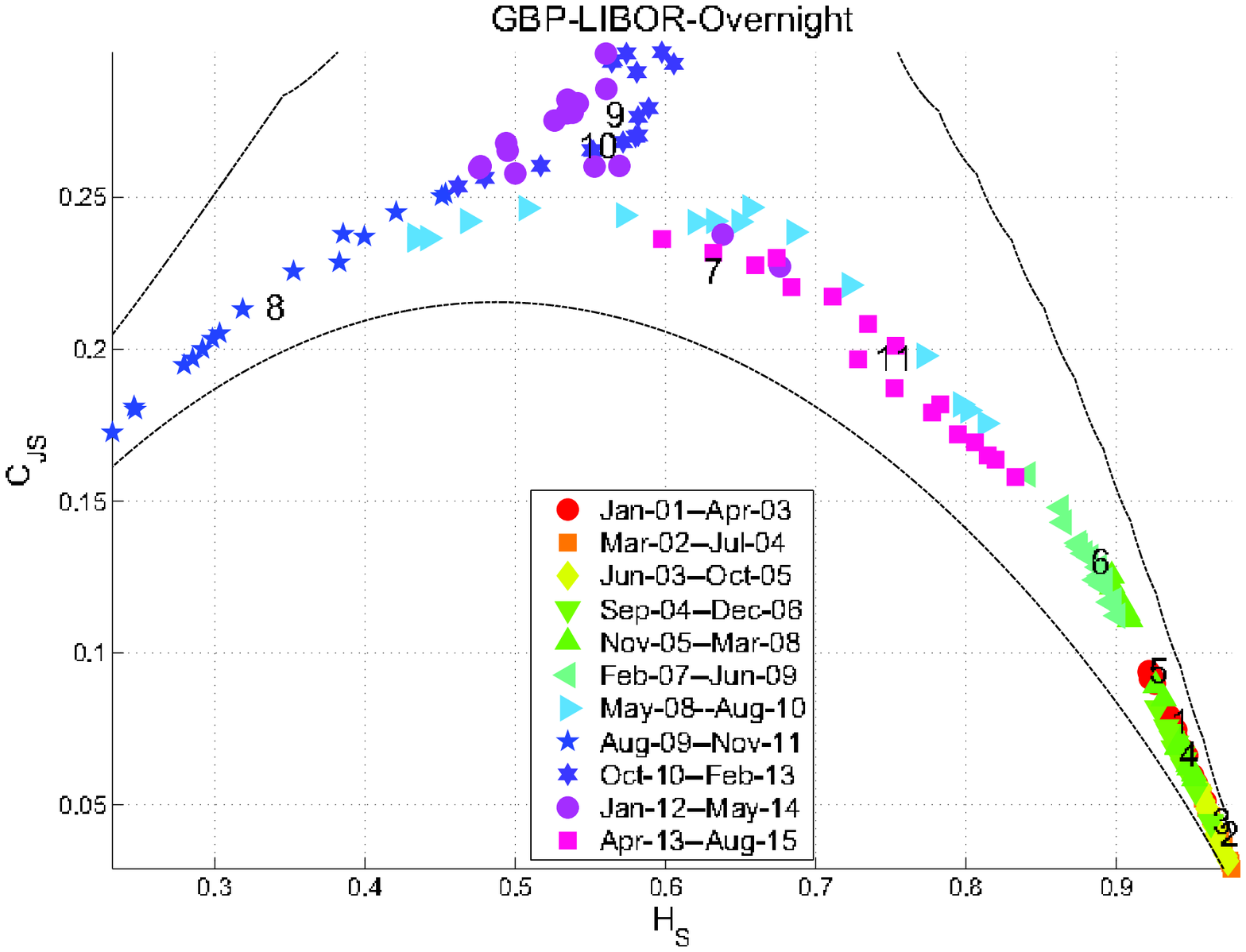}
        }%
        \subfigure{%
           \label{fig:CECPgbp1W}
           \includegraphics[width=0.45\textwidth]{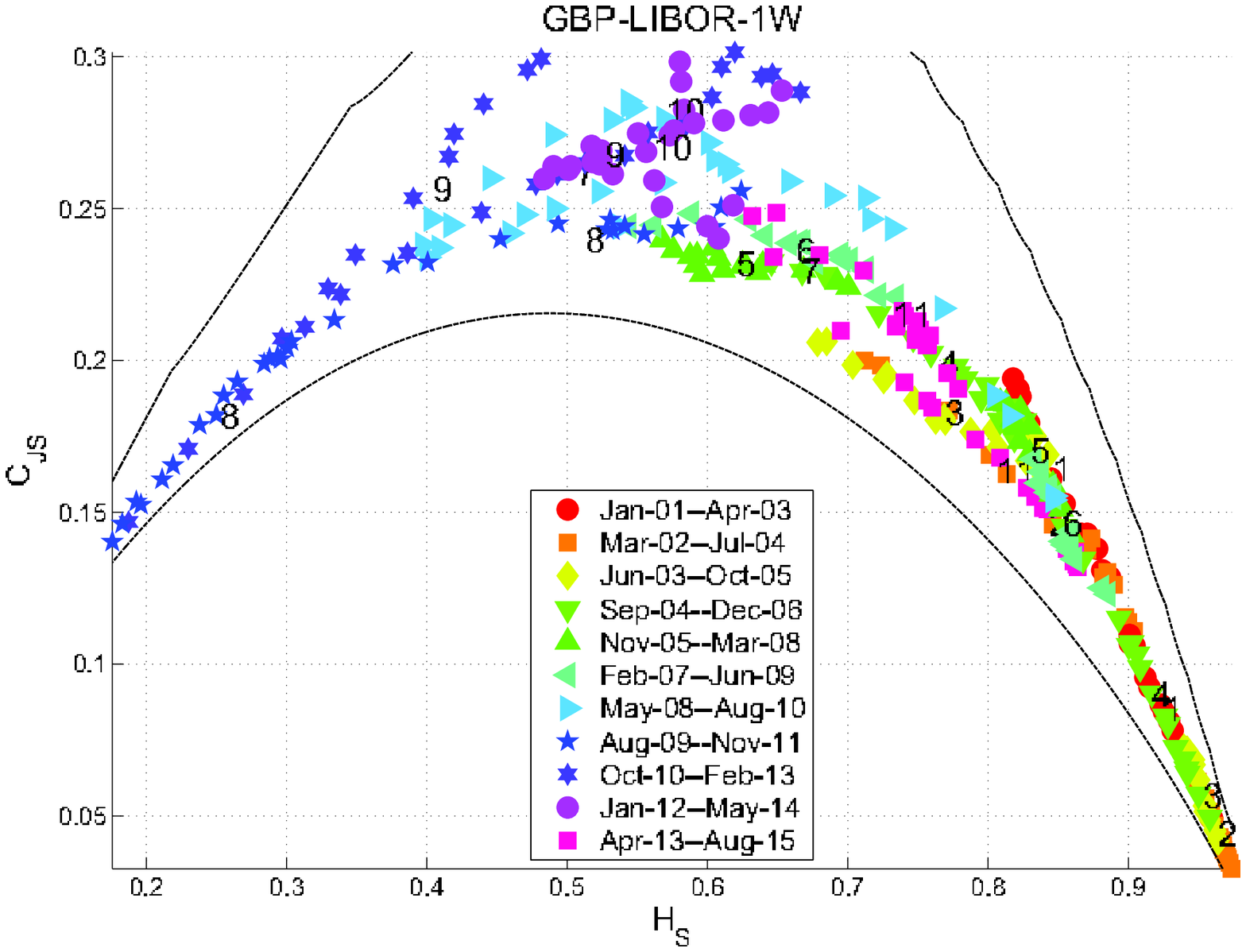}
        }
\\%  ------- End of the first row ----------------------%
        \subfigure{%
           \label{fig:CECPgbp1M}
           \includegraphics[width=0.45\textwidth]{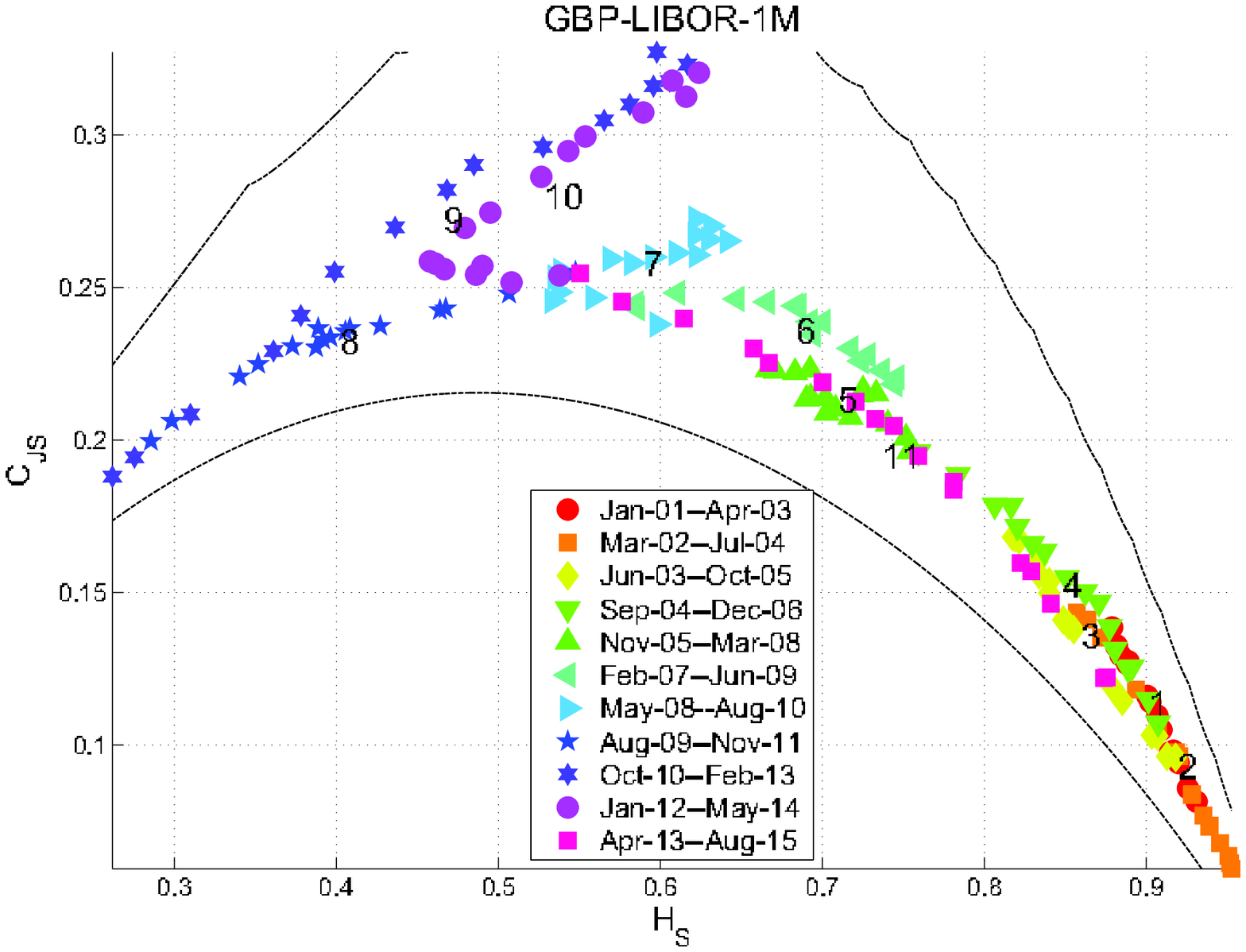}
        }
          \subfigure{%
           \label{fig:CECPgbp2M}
           \includegraphics[width=0.45\textwidth]{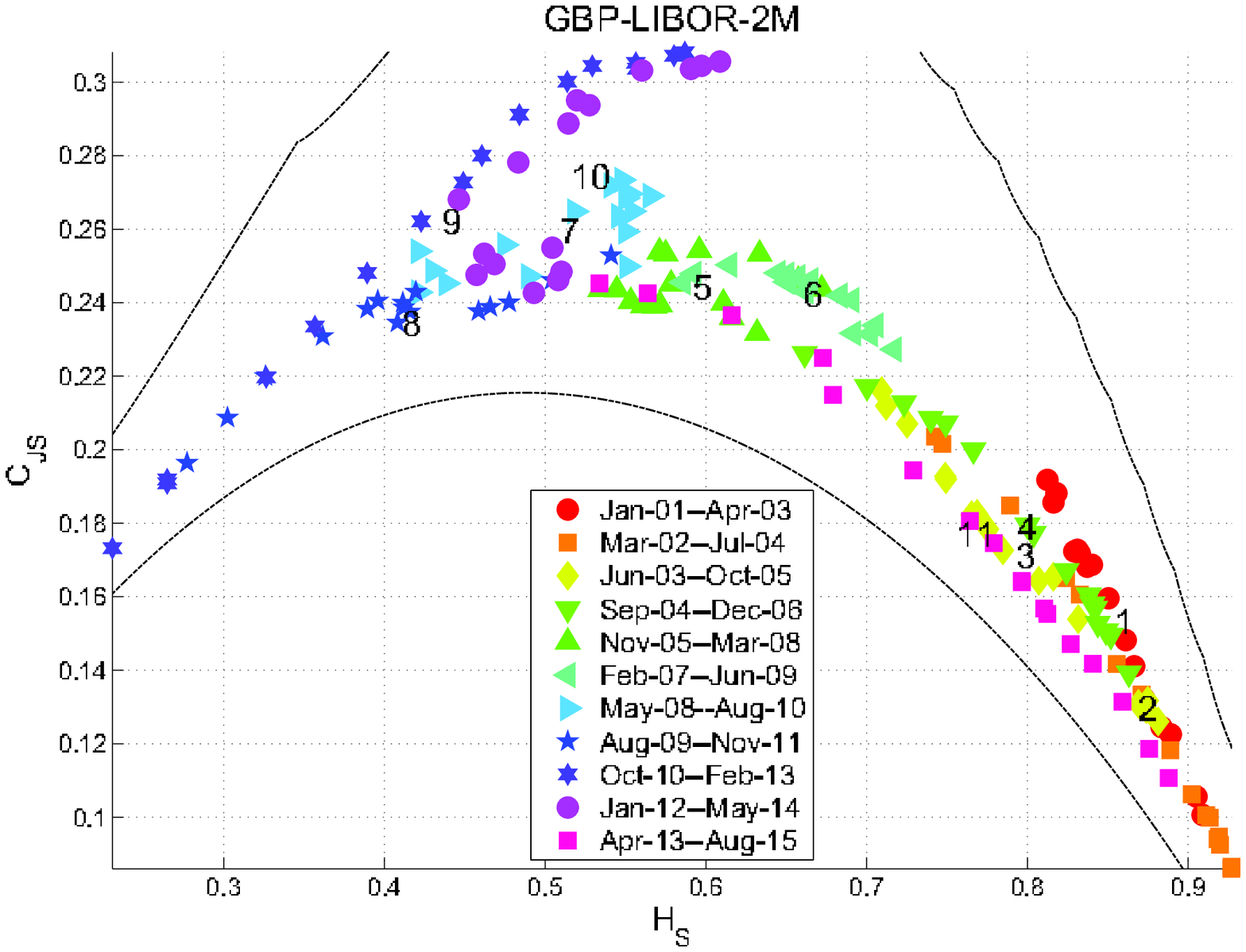}
        } 
}
\caption{Complexity Entropy Causality Plane, with $D=4,\tau=1,\delta=20$ of GBP Libor for different maturities:
         overnight (O/N), one week (1W), one month (1M), two months (2M).
         Numbers $\{1, \cdots, 11\}$ are the central points of each of the clusters. 
         The solid lines represent the upper and lower bounds of the quantifiers as computed by Mart\'{\i}n 
         {\it et al.\/} \cite{paper:martin2006}
    }%
   \label{fig:CECP_GBPlibor1}
\end{figure}

\begin{figure}[!ht]
    \centering
{
%\\%------- End of the second row ----------------------%        
       \subfigure{%
          \label{fig:CECPgbp3M}
          \includegraphics[width=0.45\textwidth]{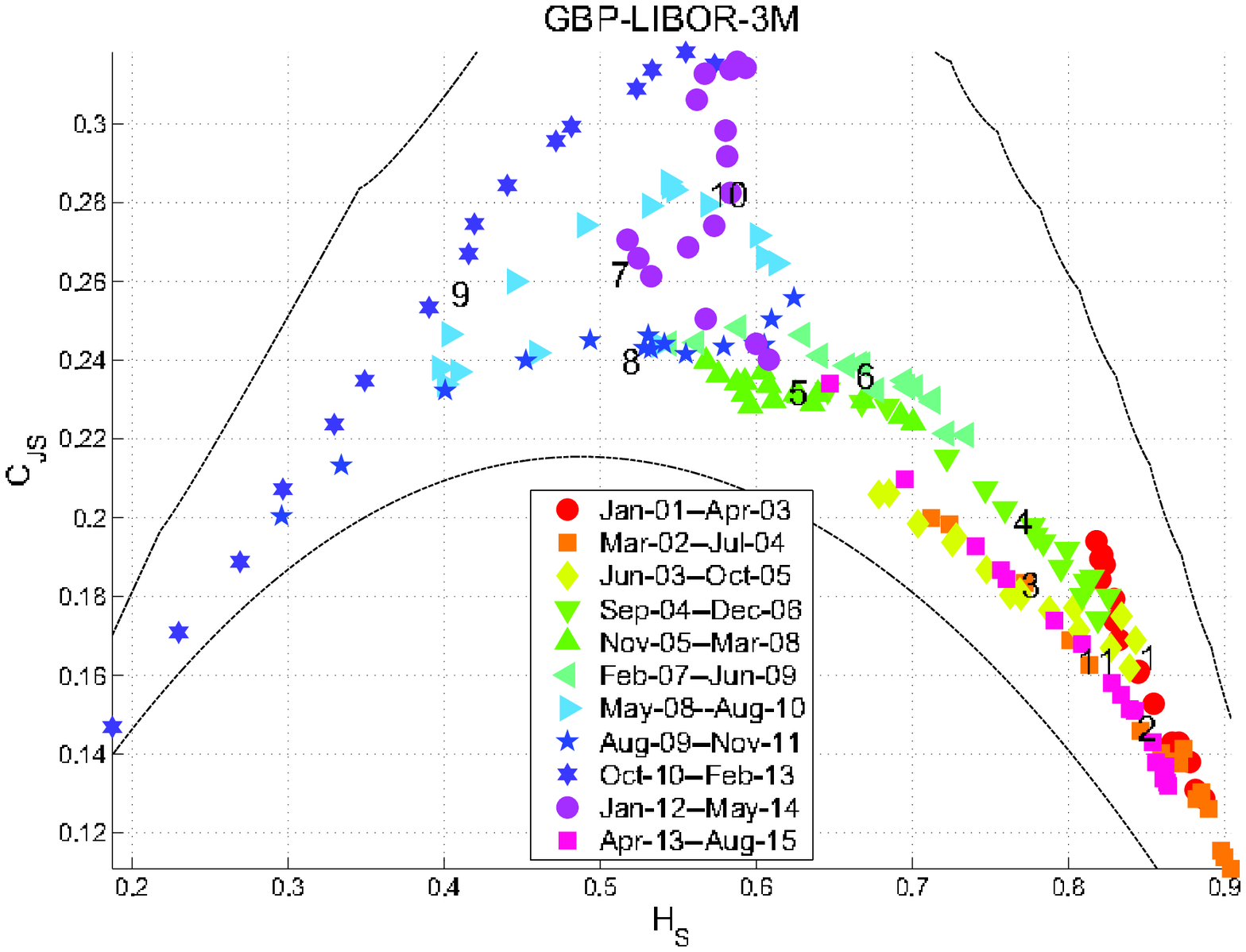}
       }
       \subfigure{%
          \label{fig:CECPgbp6M}
          \includegraphics[width=0.45\textwidth]{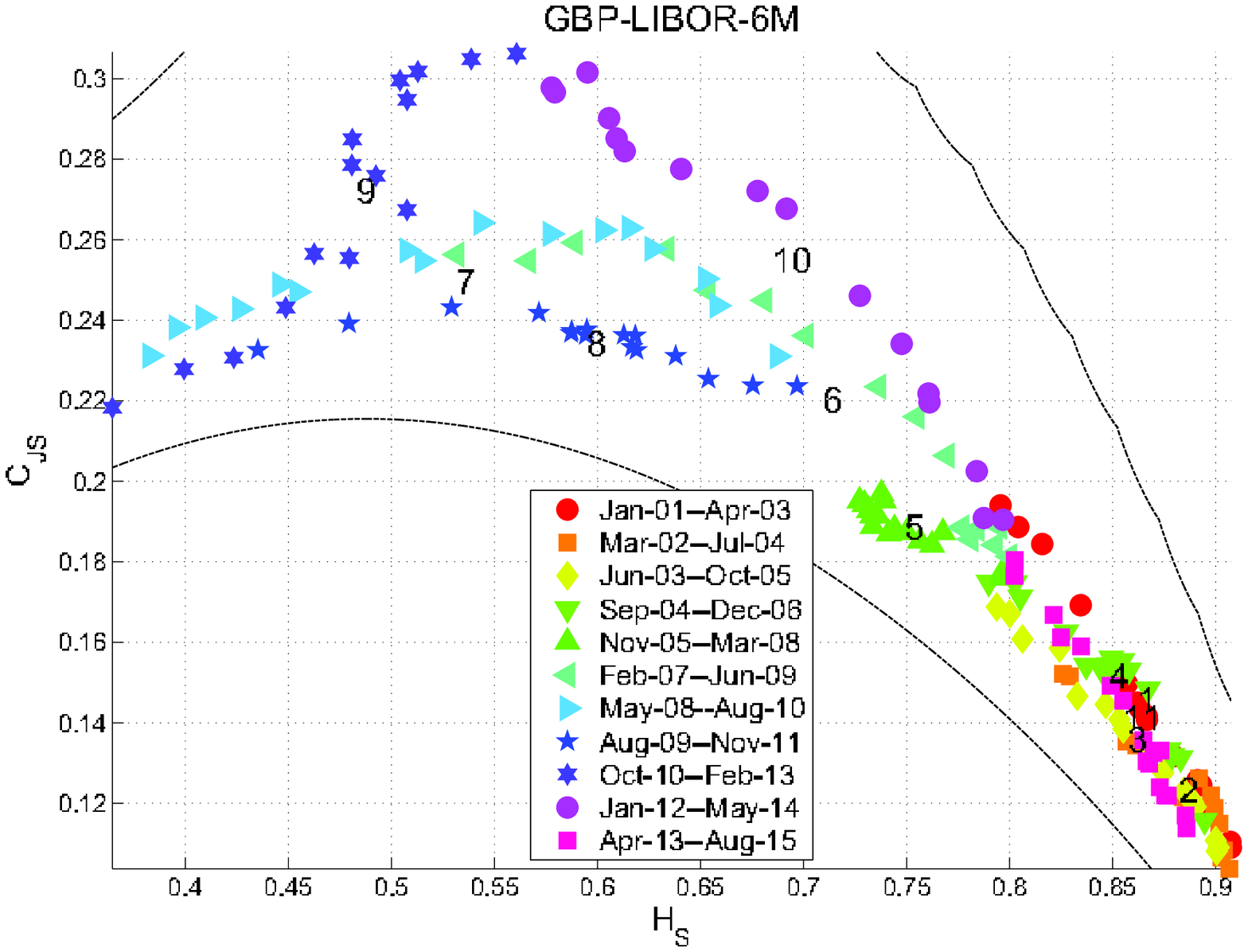}
      }
\\%------- End of the third row ----------------------%  
       \subfigure{%
          \label{fig:CECPgbp12M}
          \includegraphics[width=0.45\textwidth]{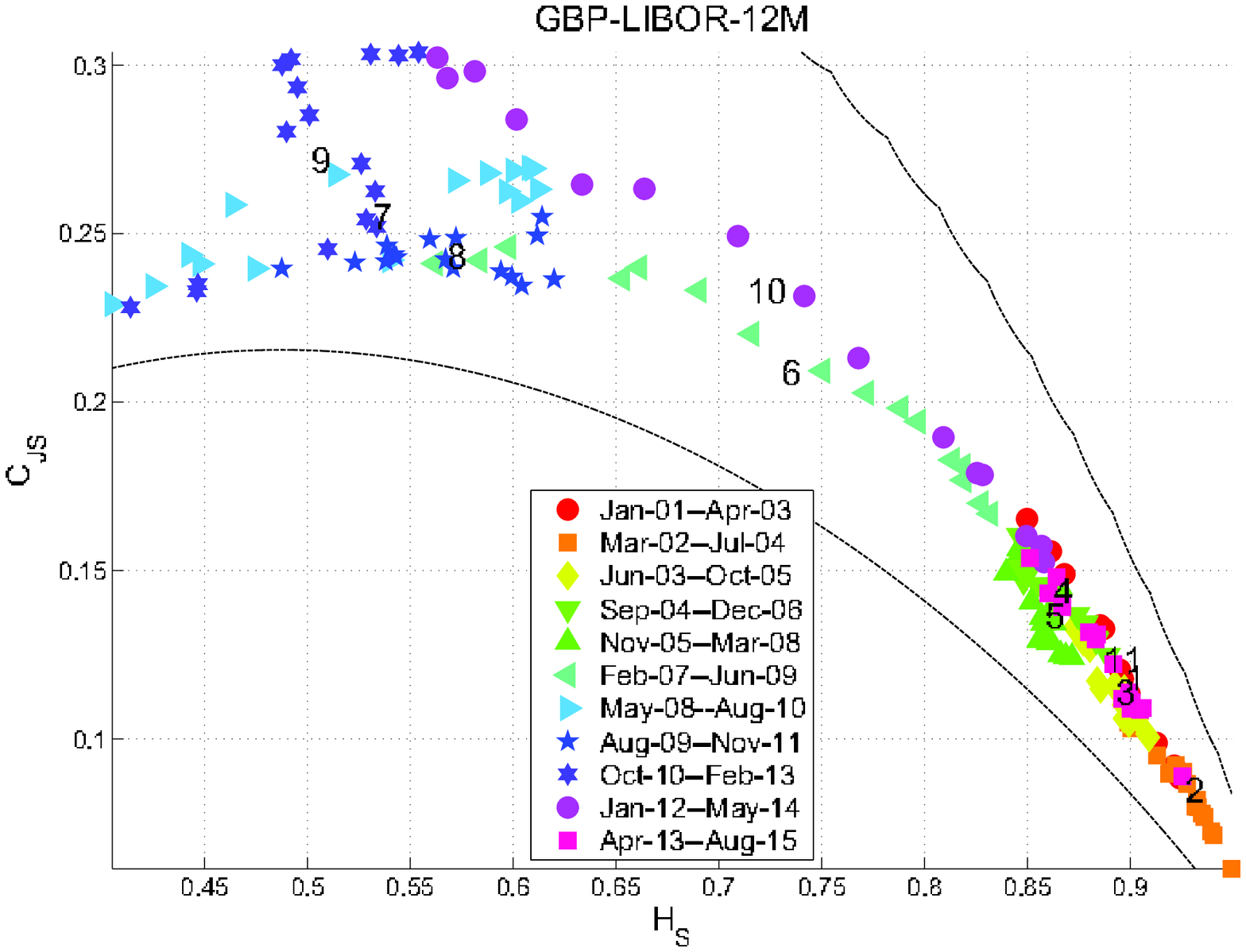}
       }
}
\caption{Complexity Entropy Causality Plane, with $D=4,\tau=1,\delta=20$ of GBP Libor for different            maturities (continuation): three months (3M), six months (6M) and twelve months (12M). 
Numbers $\{1, \cdots,11\}$ are the central points of each of the clusters. The solid lines represent the upper and lower bounds of the quantifiers as computed by Mart\'{\i}n {\it et al.\/} \cite{paper:martin2006}
    }%
   \label{fig:CECP_GBPlibor2}
\end{figure}

\begin{figure}[!ht]
    \centering
{
        \subfigure{%
            \label{fig:CECPeurON}
            \includegraphics[width=0.45\textwidth]{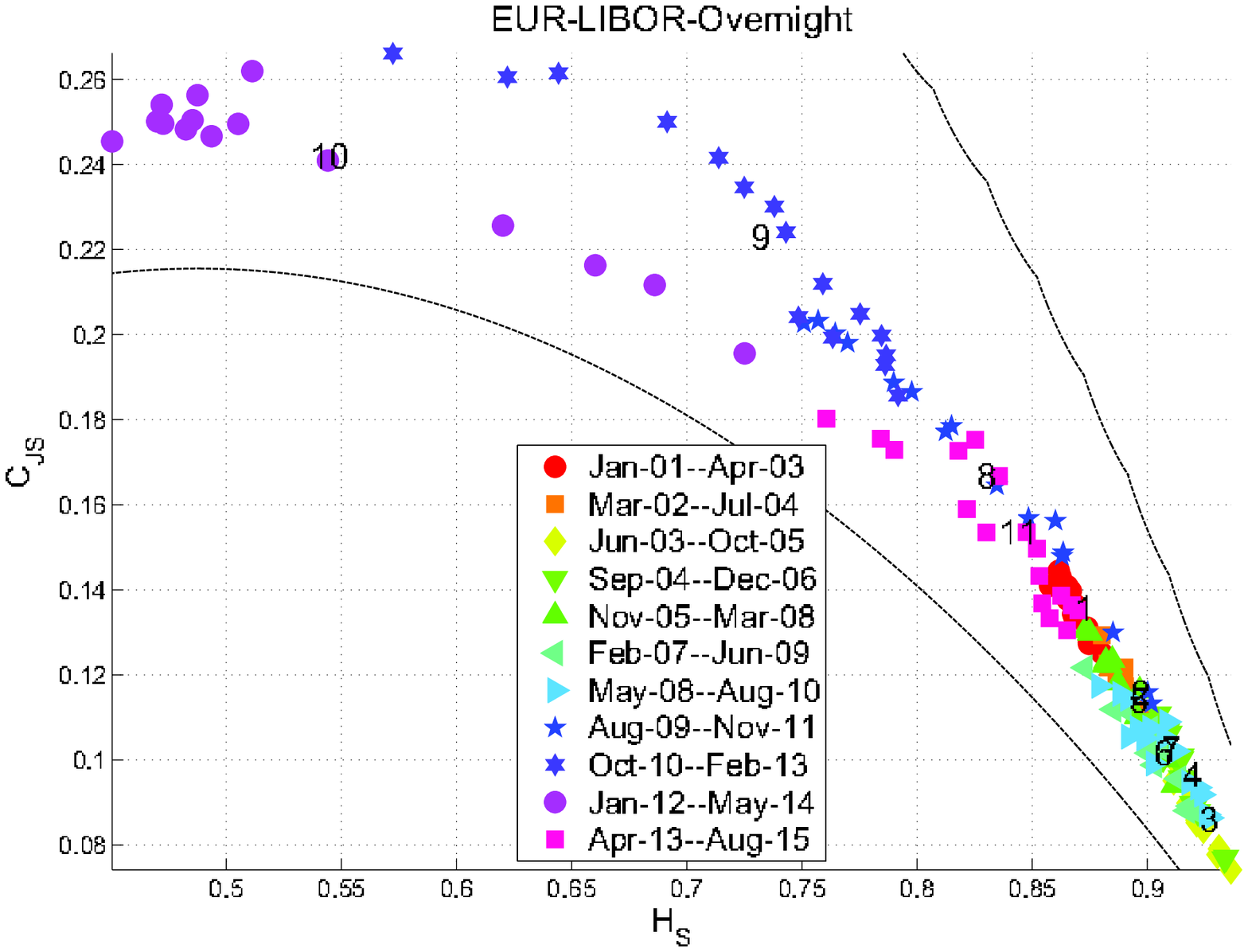}
        }%
        \subfigure{%
           \label{fig:CECPeur1W}
           \includegraphics[width=0.45\textwidth]{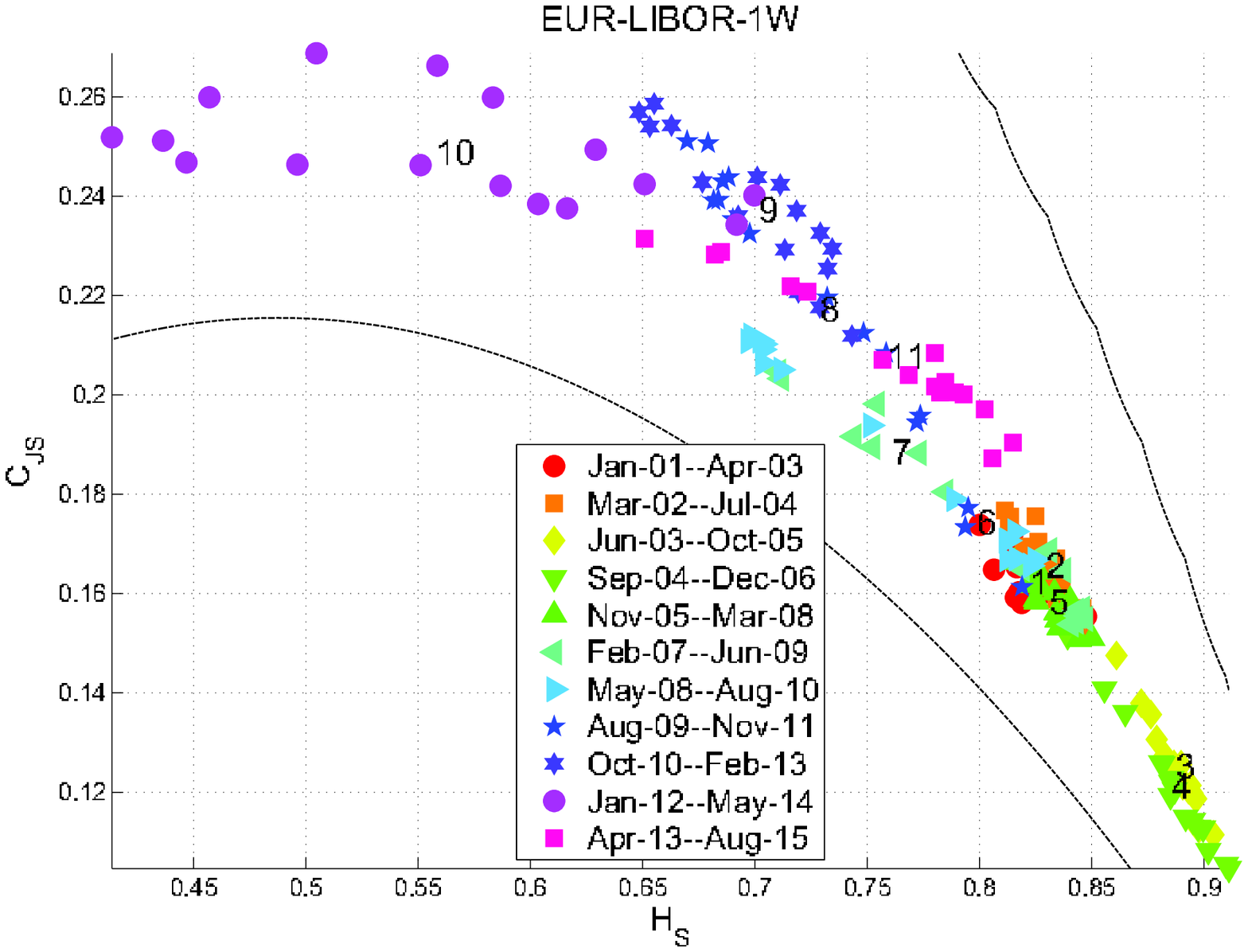}
        }
\\%  ------- End of the first row ----------------------%
        \subfigure{%
           \label{fig:CECPeur1M}
           \includegraphics[width=0.45\textwidth]{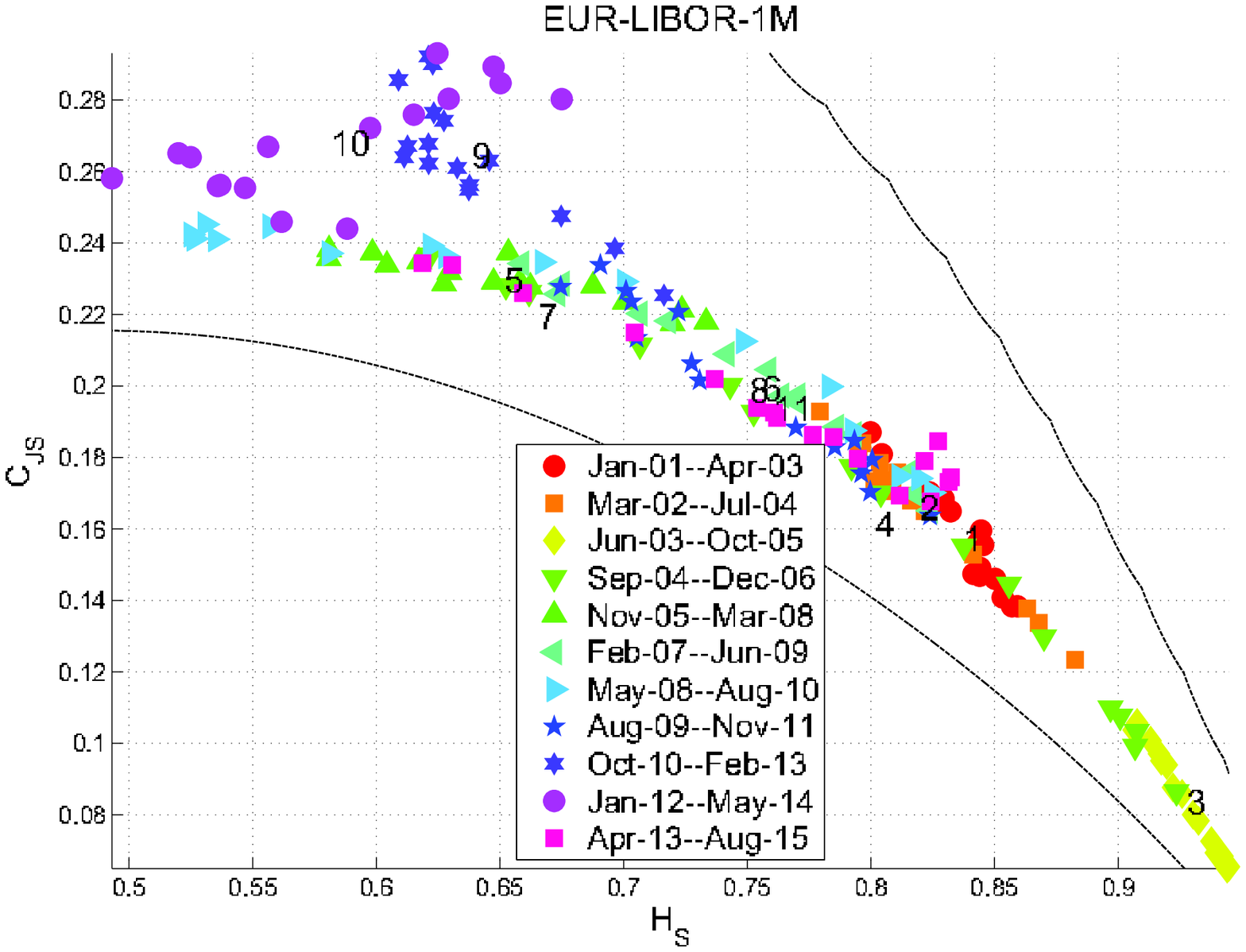}
        }
          \subfigure{%
           \label{fig:CECPeur2M}
           \includegraphics[width=0.45\textwidth]{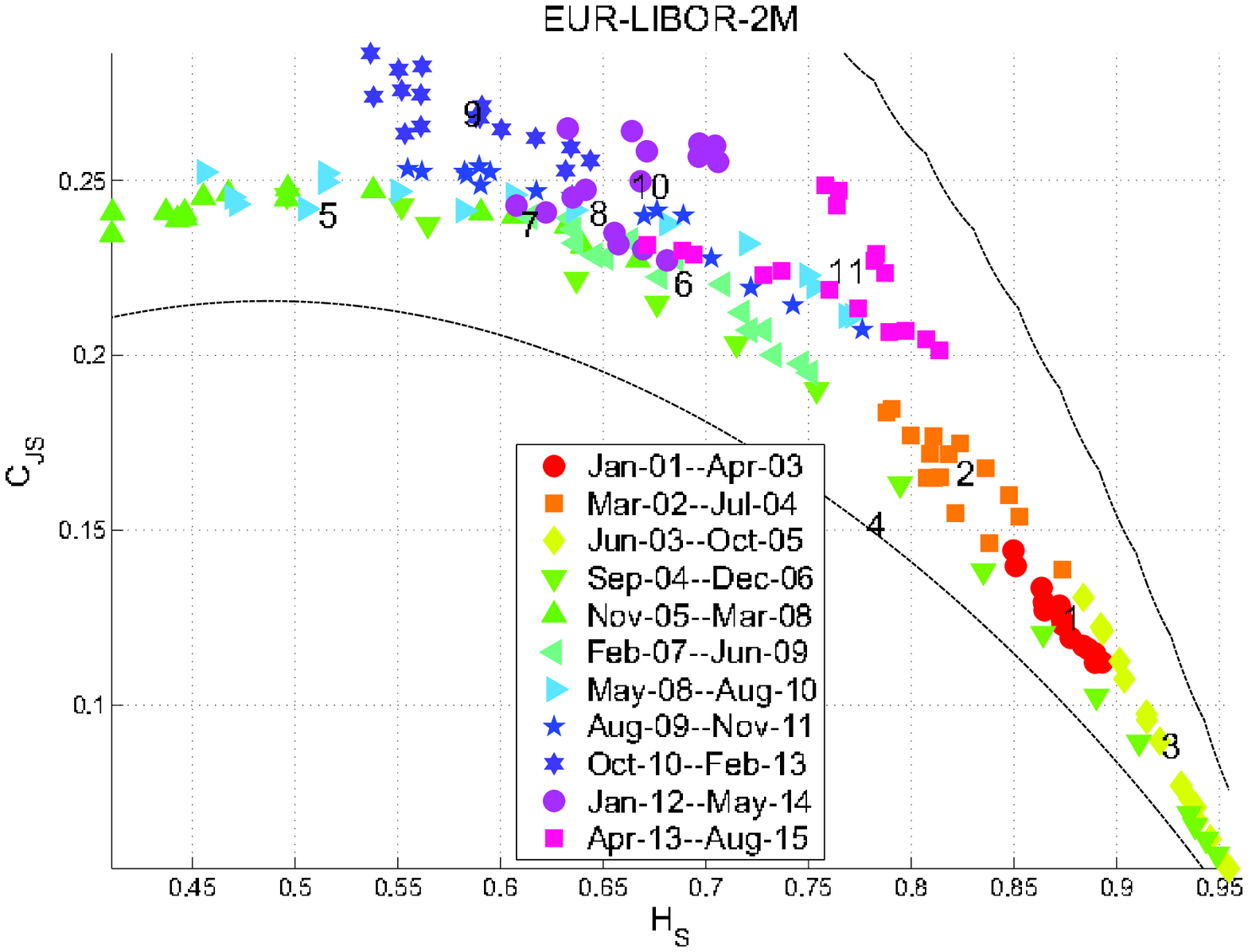}
        }%
%\\%------- End of the second row ----------------------% 
}
    \caption{Complexity Entropy Causality Plane, with $D=4,\tau=1,\delta=20$ of EUR Libor for different maturities: overnight (O/N), one week (1W), one month (1M), two months (2M).
Numbers $\{1, \cdots, 11\}$ are the central points of each of the clusters. 
The solid lines represent the upper and lower bounds of the quantifiers as computed by Mart\'{\i}n {\it et al.\/} \cite{paper:martin2006}
    }%
   \label{fig:CECP_EURlibor1}
\end{figure}

\begin{figure}[!ht]
    \centering
{      
       \subfigure{%
          \label{fig:CECPeur3M}
          \includegraphics[width=0.45\textwidth]{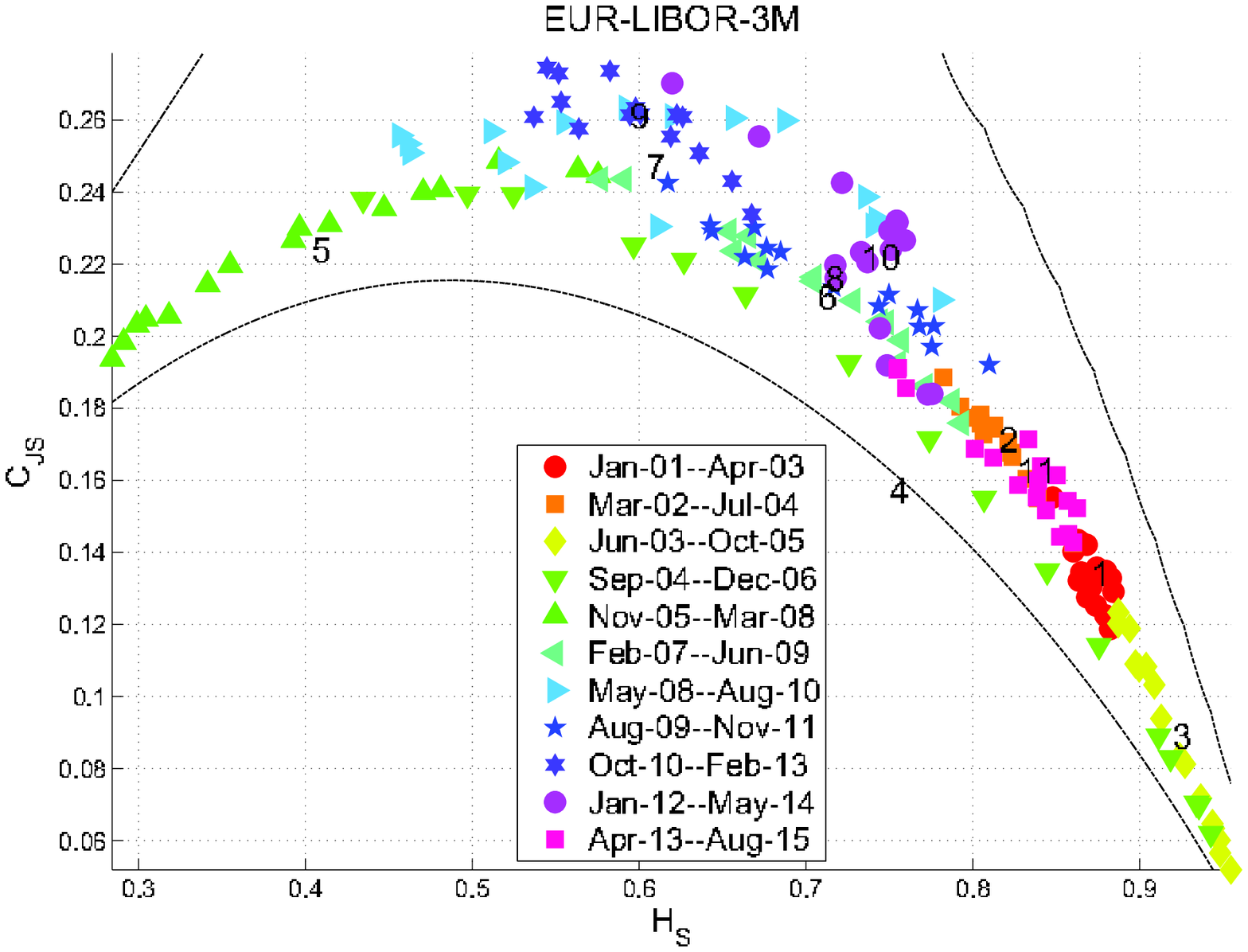}
       }
       \subfigure{%
          \label{fig:CECPeur6M}
          \includegraphics[width=0.45\textwidth]{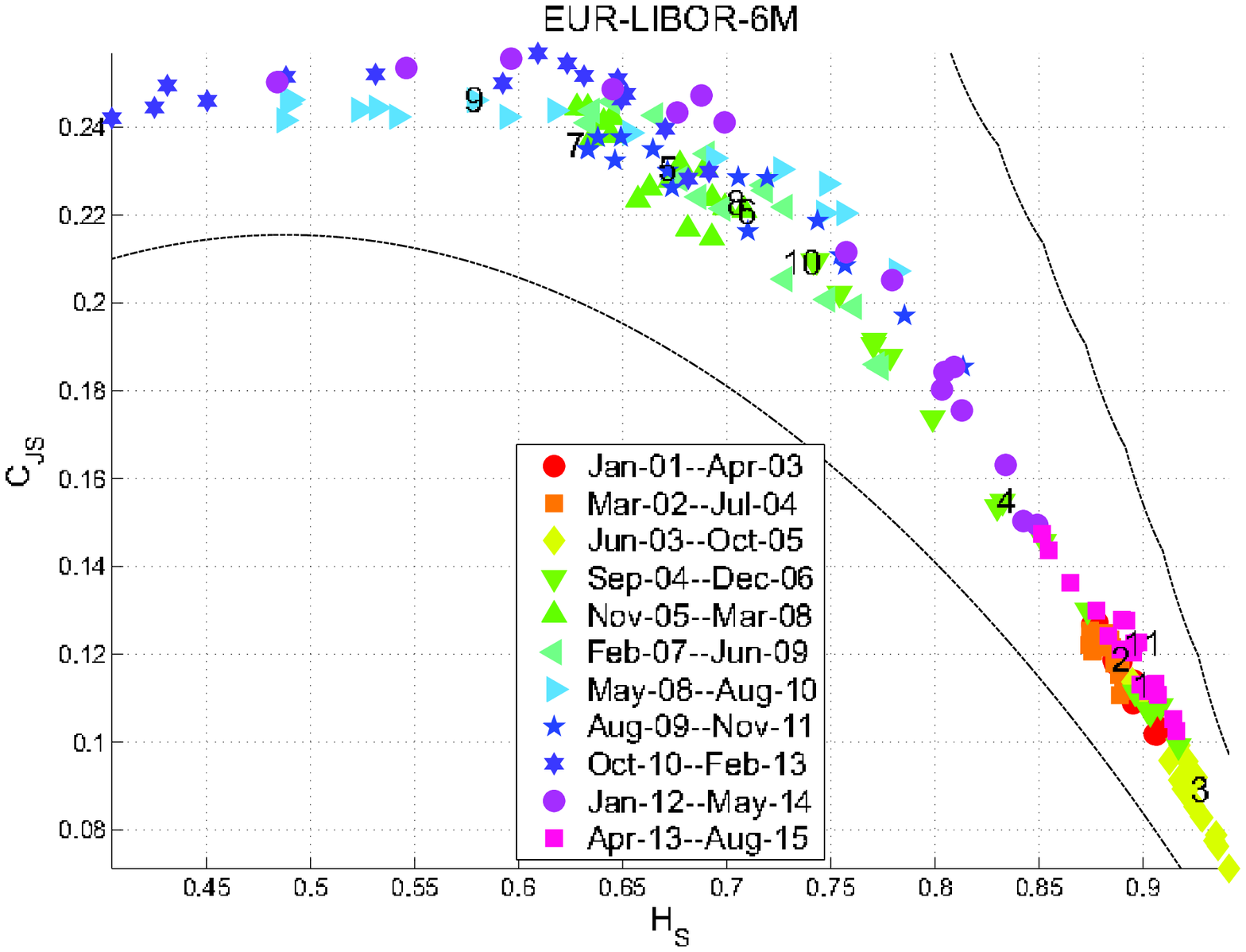}
      }
\\%------- End of the third row ----------------------%  
       \subfigure{%
          \label{fig:CECPeur12M}
          \includegraphics[width=0.45\textwidth]{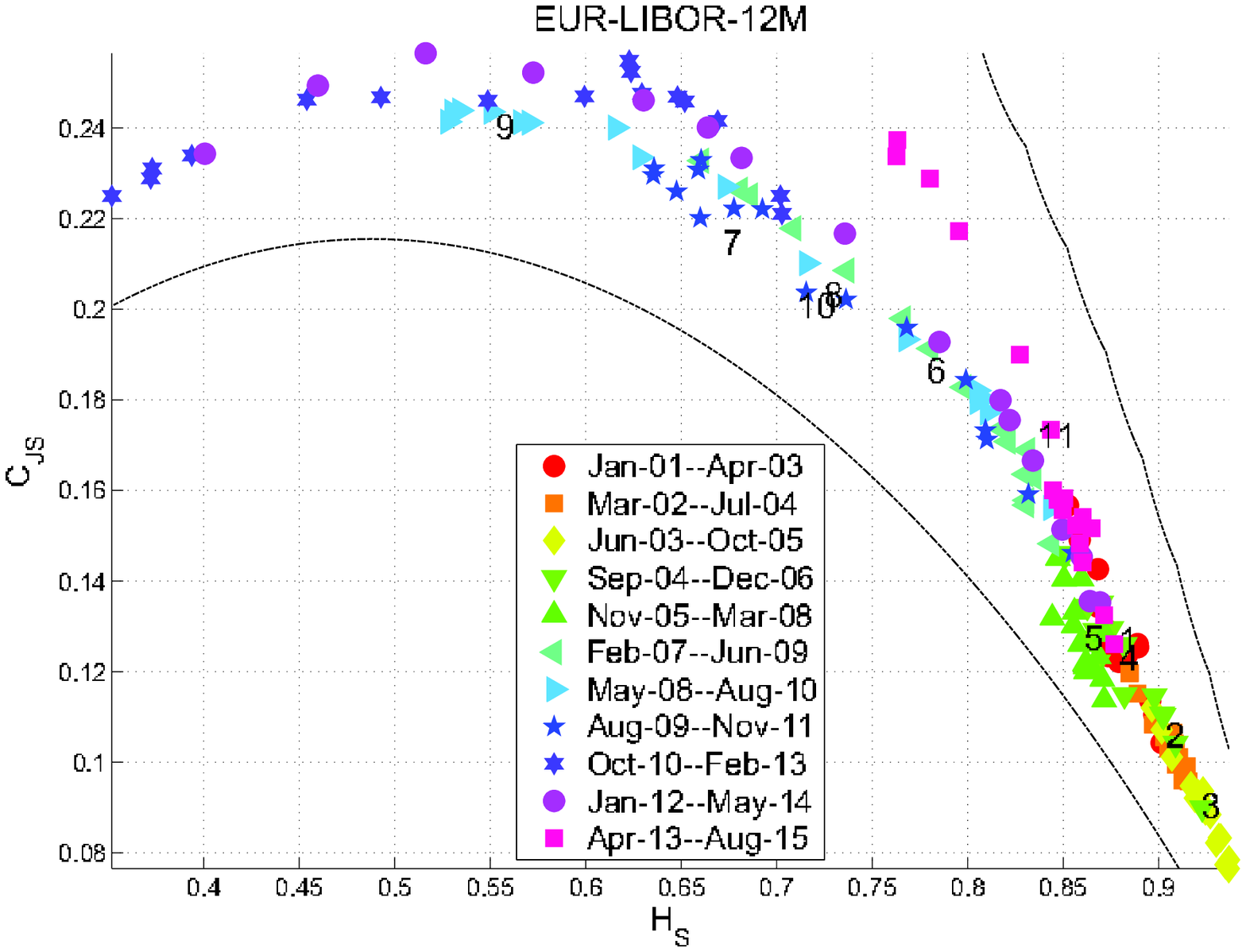}
       }
}
    \caption{Complexity Entropy Causality Plane, with $D=4,\tau=1,\delta=20$ of EUR Libor for different maturities (continuation): three months (3M), six months (6M) and twelve months (12M). 
Numbers $\{1 , \cdots ,11 \}$ are the central points of each of the clusters. The solid lines represent the upper and lower bounds of the quantifiers as computed by Mart\'{\i}n {\it et al.\/} \cite{paper:martin2006}
    }%
   \label{fig:CECP_EURlibor2}
\end{figure}

\begin{figure}[!ht]
    \centering
{
        \subfigure{%
            \label{fig:CECPchfON}
            \includegraphics[width=0.45\textwidth]{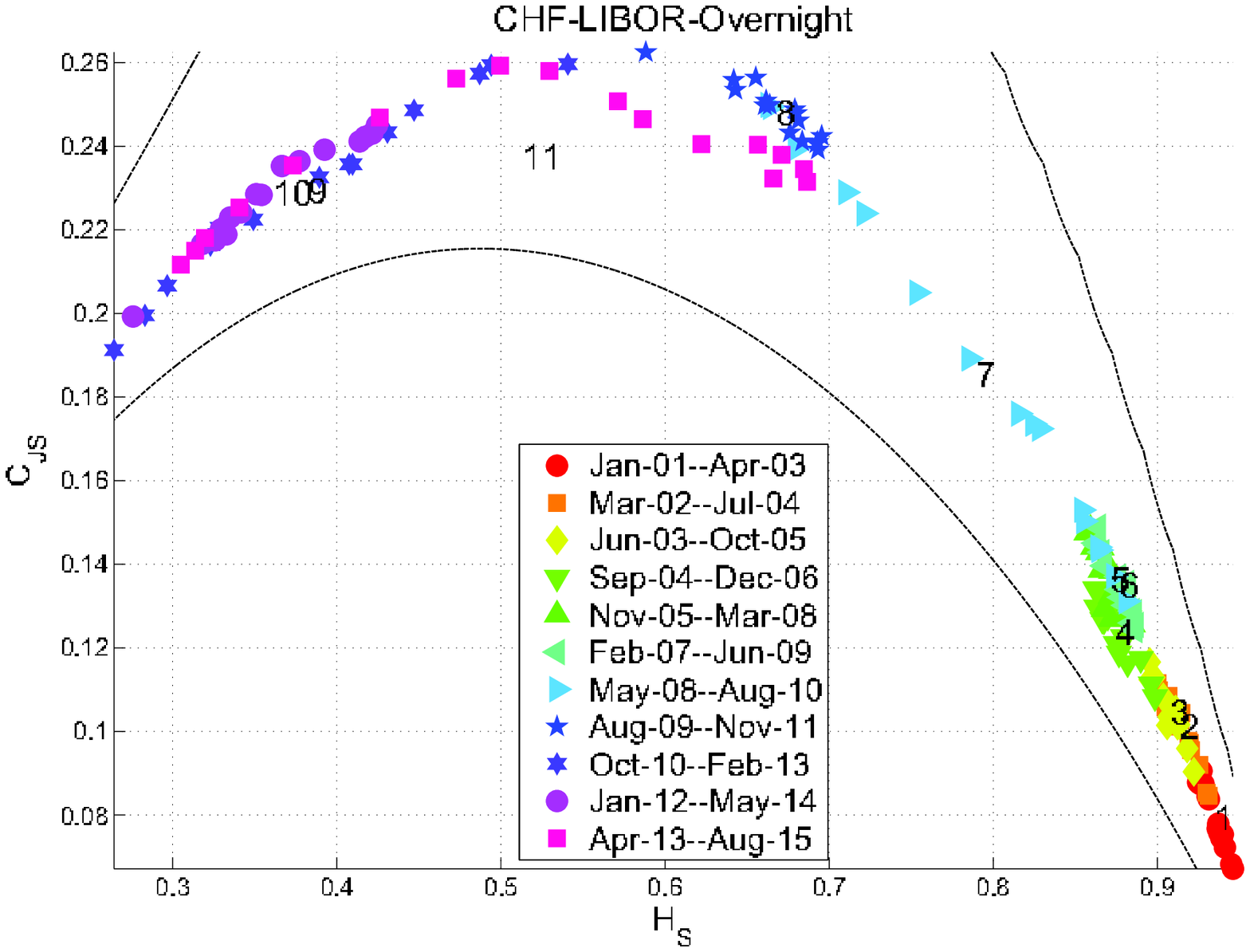}
        }%
        \subfigure{%
           \label{fig:CECPchf1W}
           \includegraphics[width=0.45\textwidth]{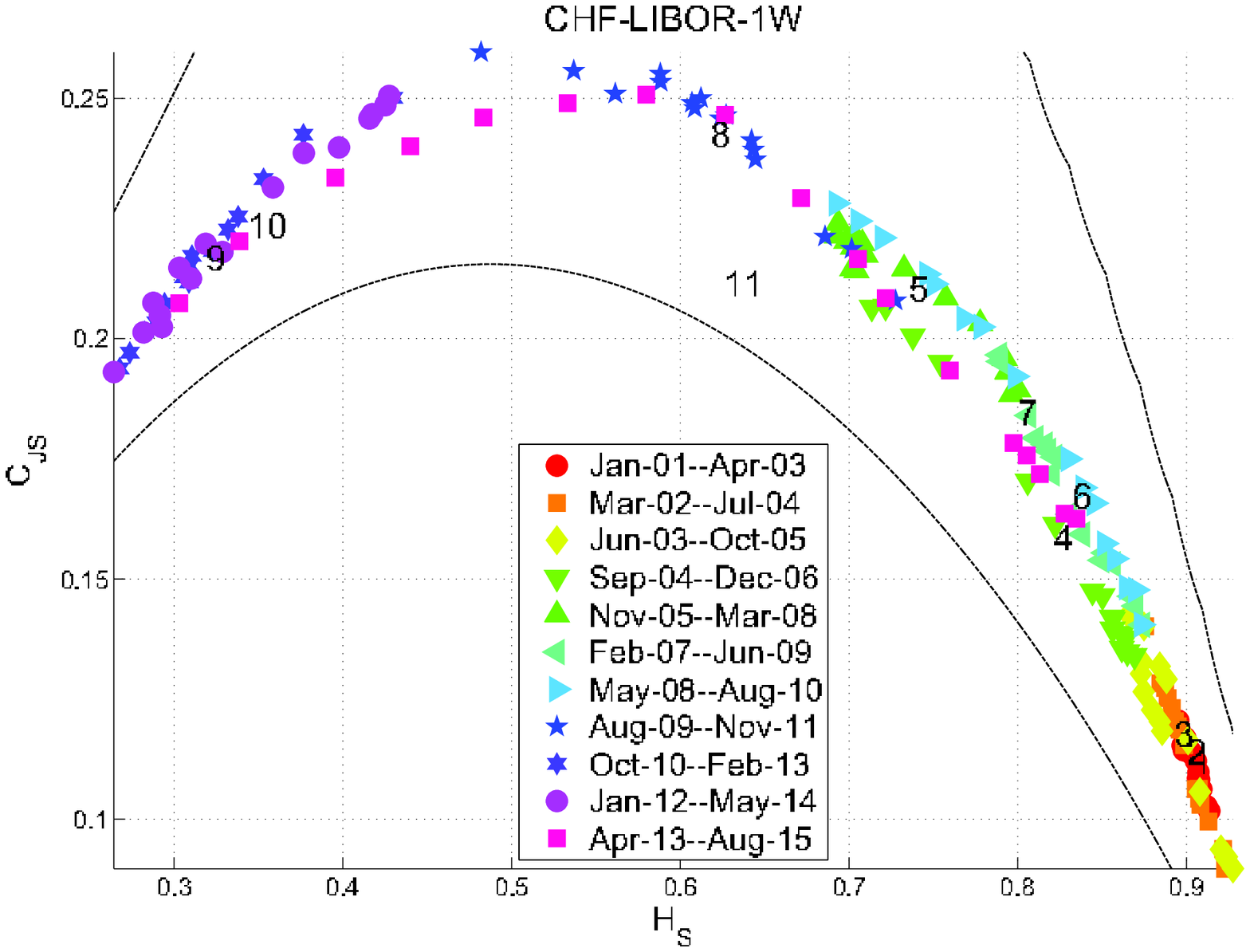}
        }
\\%  ------- End of the first row ----------------------%
        \subfigure{%
           \label{fig:CECPchf1M}
           \includegraphics[width=0.45\textwidth]{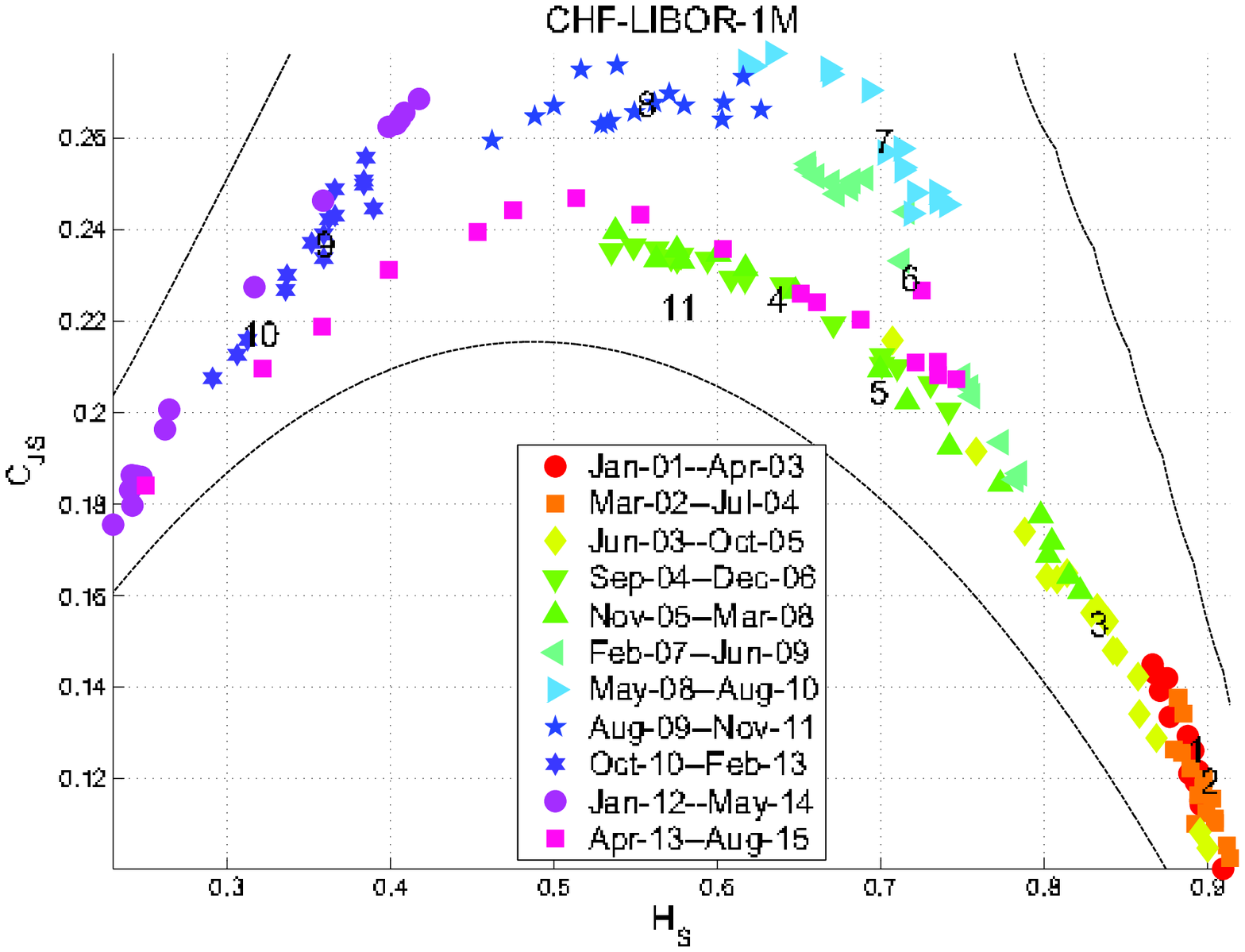}
        }
          \subfigure{%
           \label{fig:CECPchf2M}
           \includegraphics[width=0.45\textwidth]{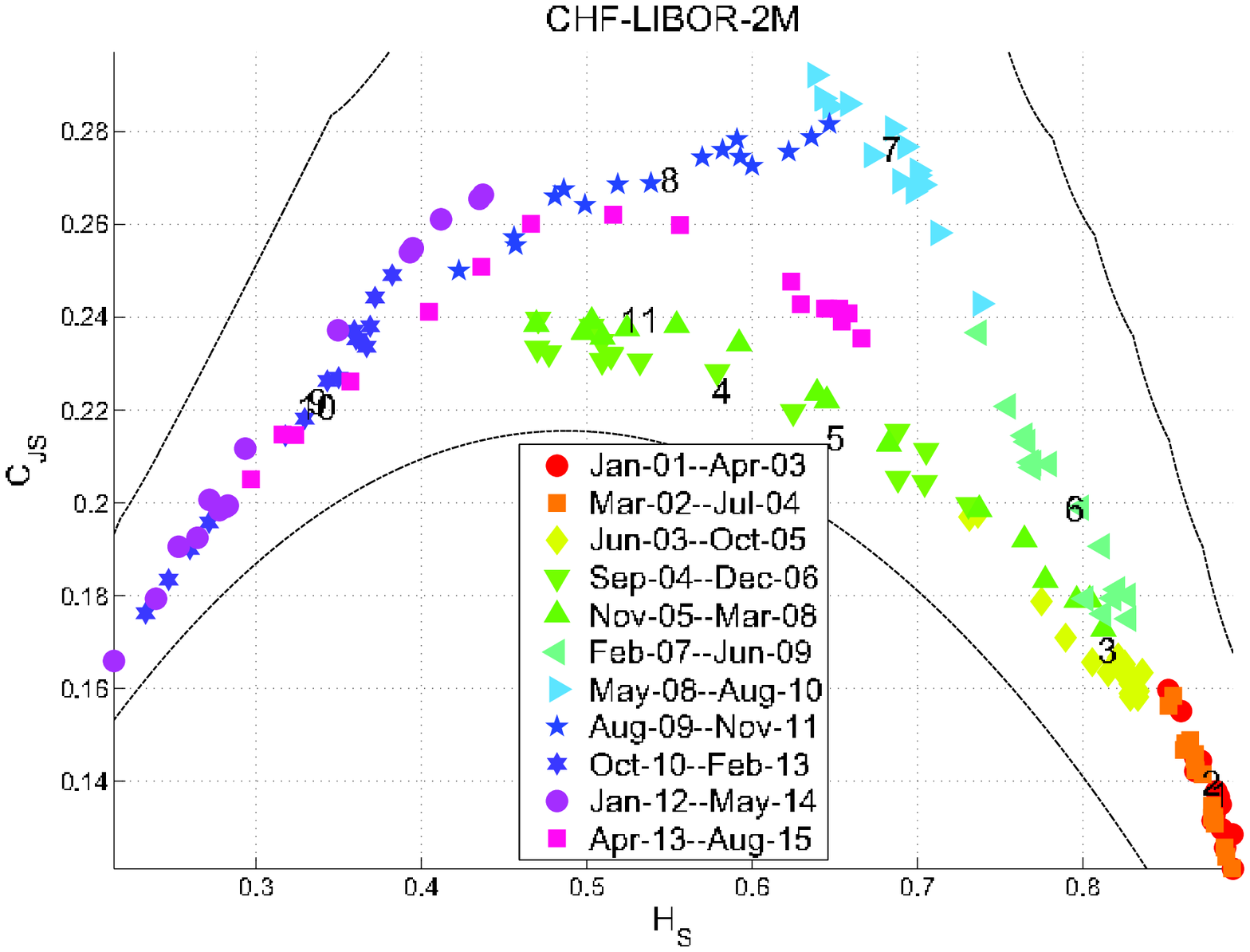}
%        }%
%\\%------- End of the second row ----------------------%        
       }
}
    \caption{Complexity Entropy Causality Plane, with $D=4,\tau=1,\delta=20$ of CHF Libor for different maturities:
overnight (O/N), one week (1W), one month (1M), two months (2M).
Numbers $\{1, \cdots, 11\}$ are the central points of each of the clusters. The solid lines represent the upper and lower bounds of the quantifiers as computed by Mart\'{\i}n {\it et al.\/} \cite{paper:martin2006}}
\label{fig:CECP_CHFlibor1}
\end{figure}

\begin{figure}[!ht]
    \centering
{
       \subfigure{%
          \label{fig:CECPchf3M}
          \includegraphics[width=0.45\textwidth]{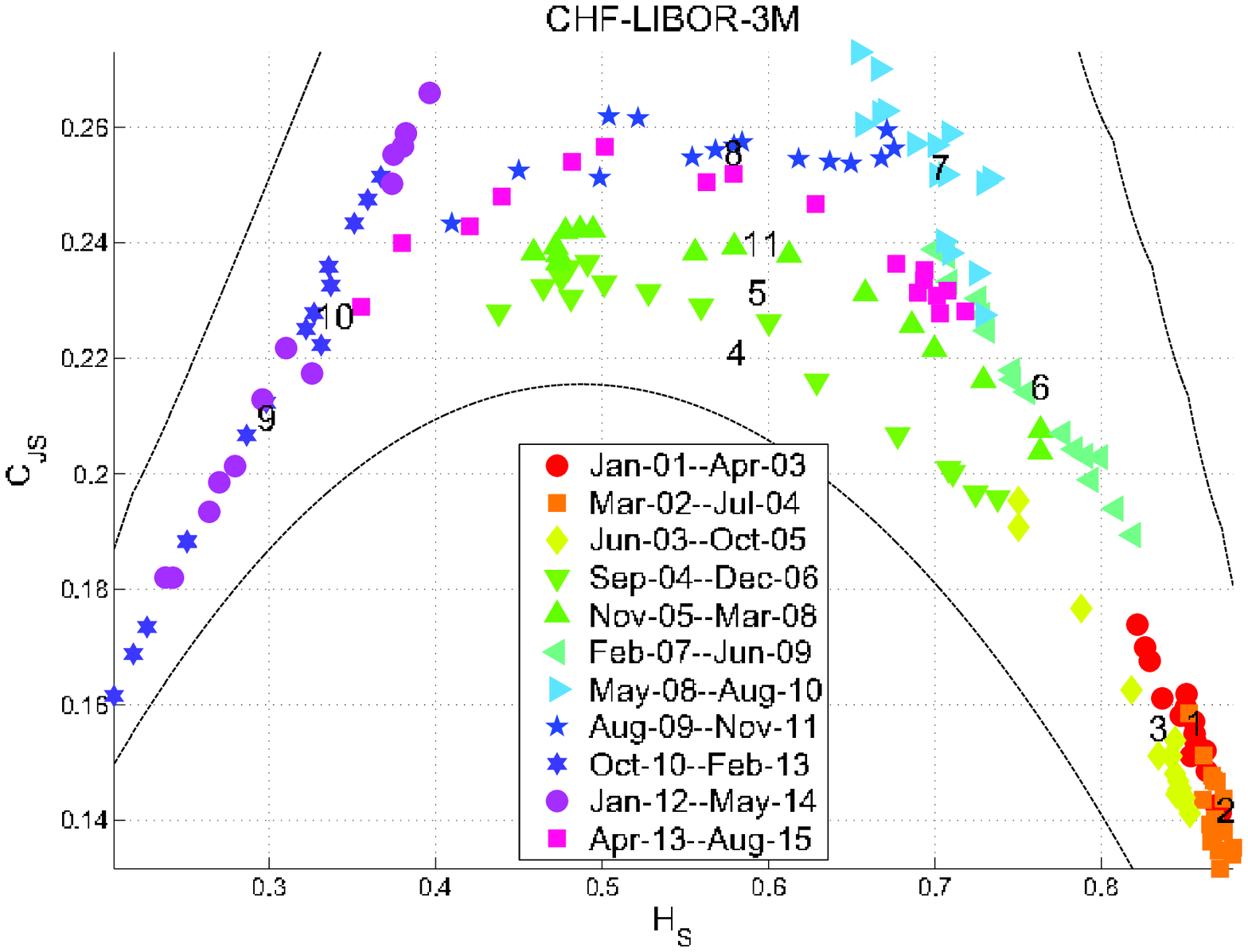}
       }
       \subfigure{%
          \label{fig:CECPchf6M}
          \includegraphics[width=0.45\textwidth]{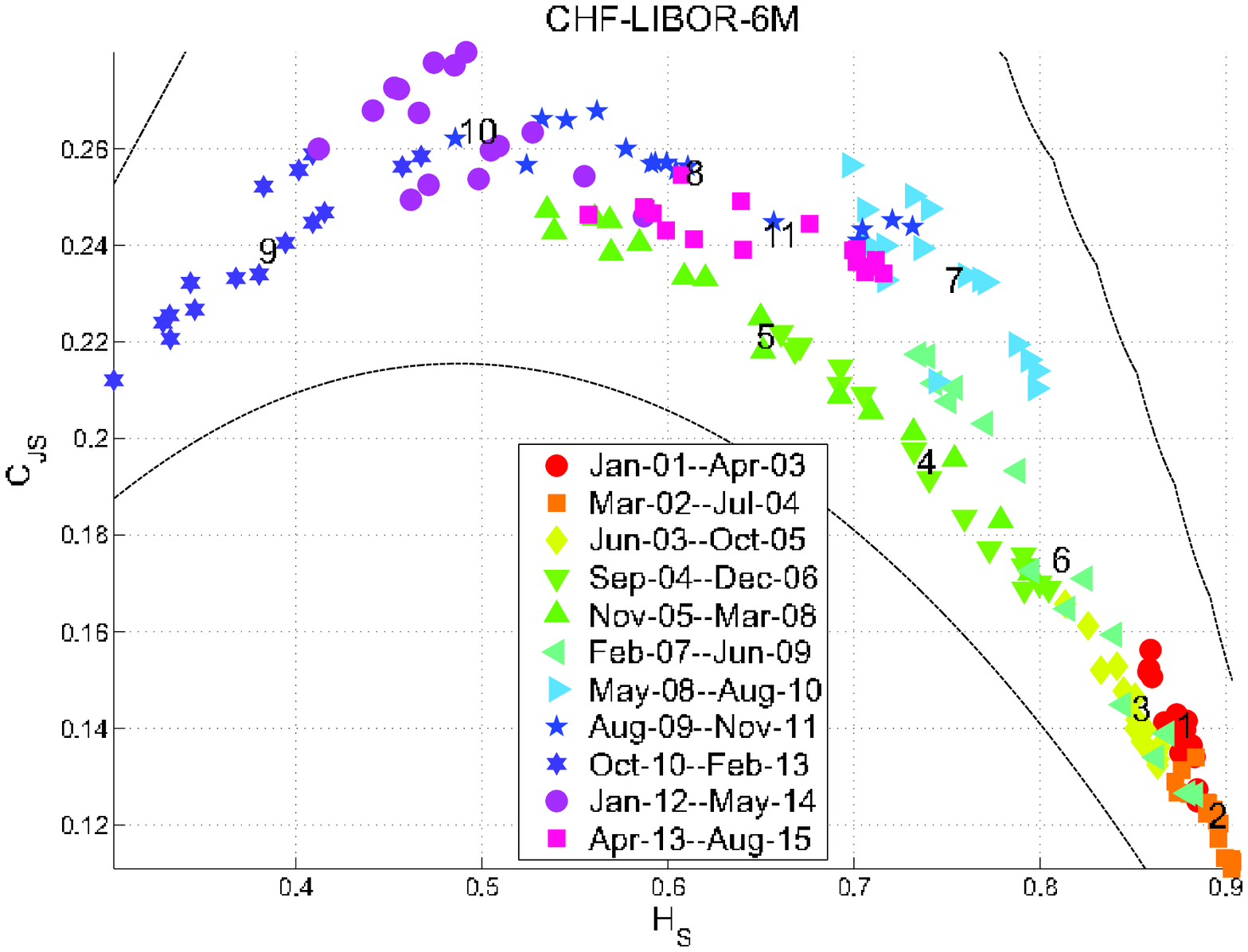}
      }
\\%------- End of the third row ----------------------%  
       \subfigure{%
          \label{fig:CECPchf12M}
          \includegraphics[width=0.45\textwidth]{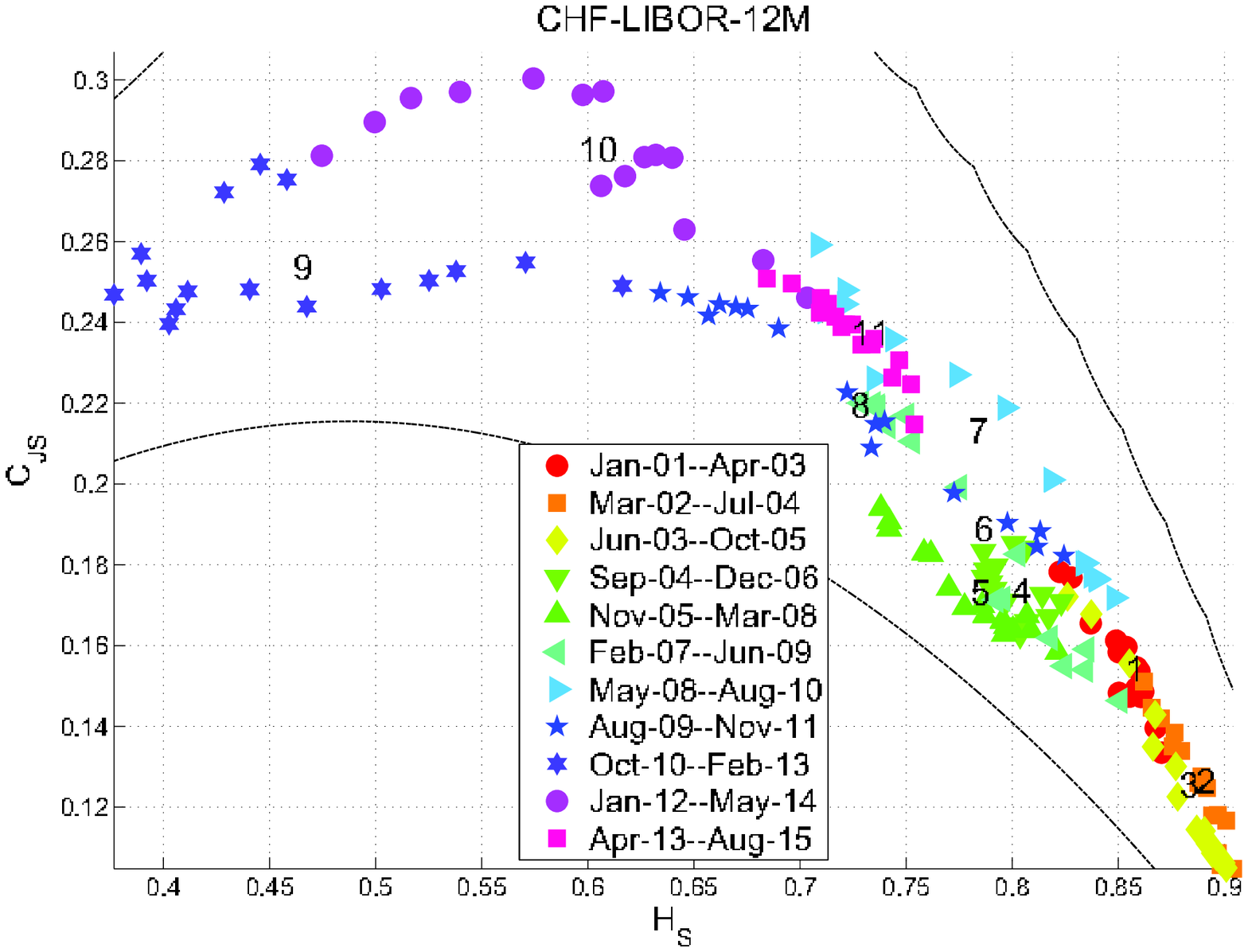}
       }
}
    \caption{Complexity Entropy Causality Plane, with $D=4,\tau=1,\delta=20$ of CHF Libor for different maturities (continuation): three months (3M), six months (6M) and twelve months (12M).
Numbers $\{1, \dots,11\}$ are the central points of each of the clusters. The solid lines represent the upper and lower bounds of the quantifiers as computed by Mart\'{\i}n {\it et al.\/} \cite{paper:martin2006}}
\label{fig:CECP_CHFlibor2}
\end{figure}

\begin{figure}[!ht]
    \centering
{
        \subfigure{%
            \label{fig:CECPjpyON}
            \includegraphics[width=0.45\textwidth]{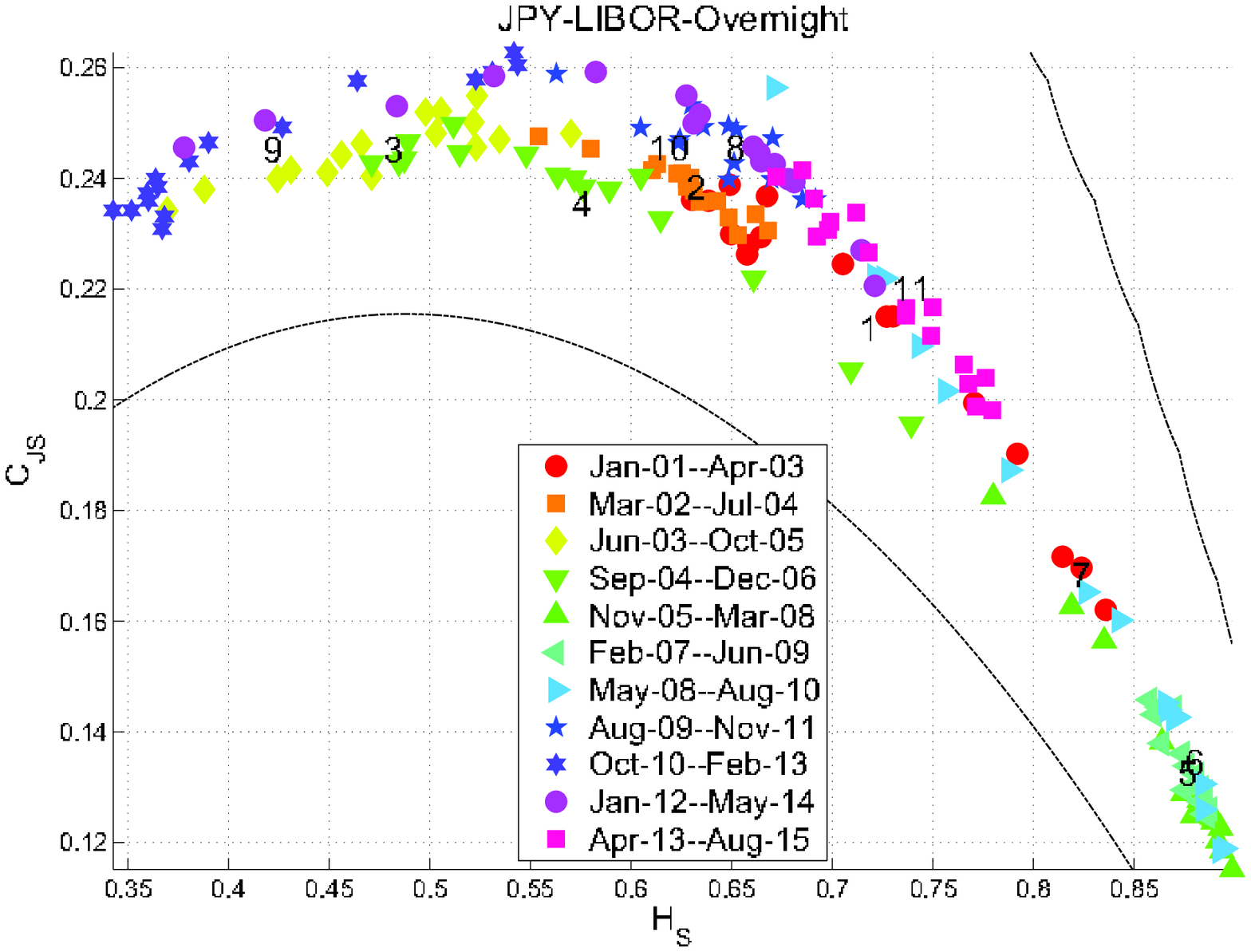}
        }%
        \subfigure{%
           \label{fig:CECPjpy1W}
           \includegraphics[width=0.45\textwidth]{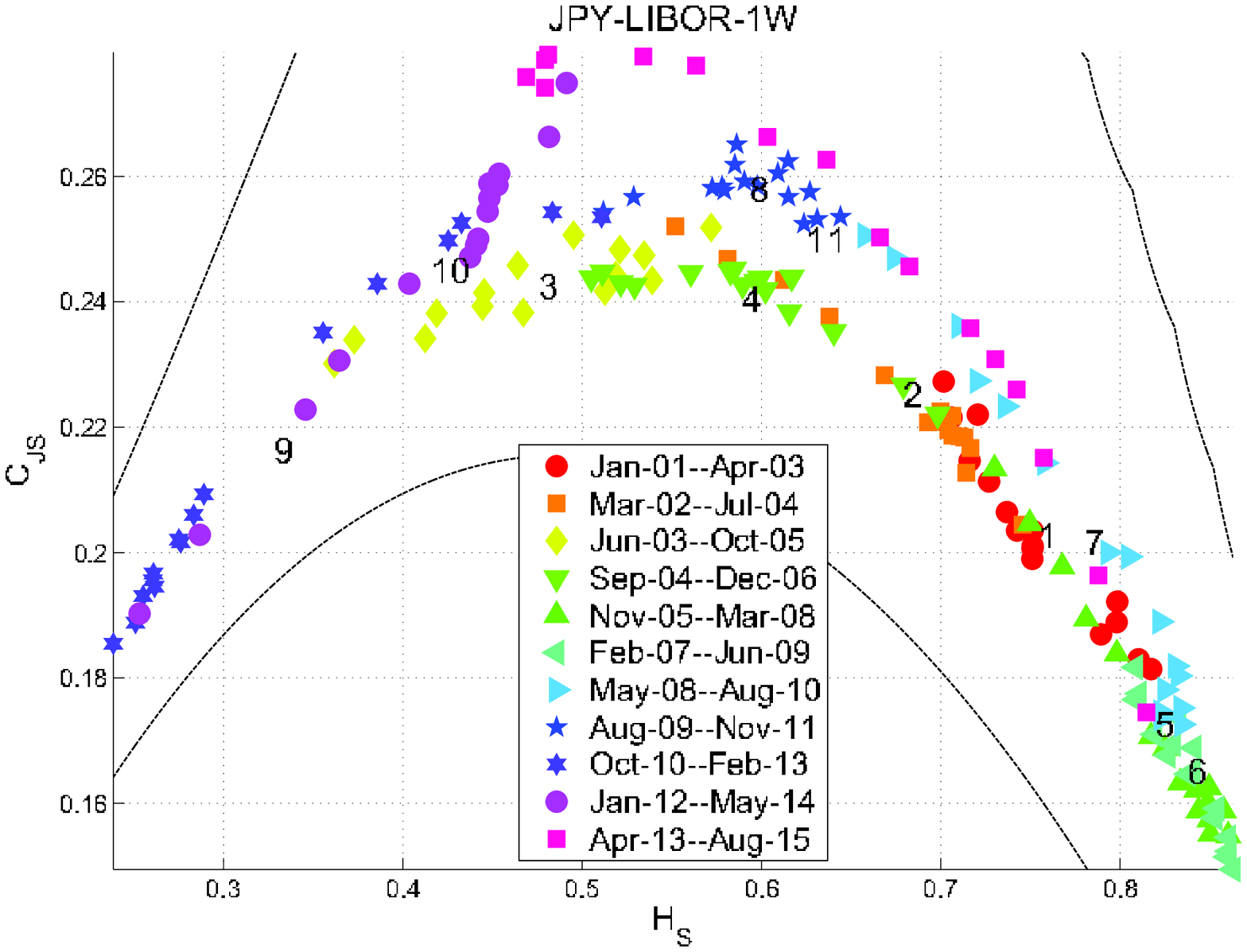}
        }
\\%  ------- End of the first row ----------------------%
        \subfigure{%
           \label{fig:CECPjpy1M}
           \includegraphics[width=0.45\textwidth]{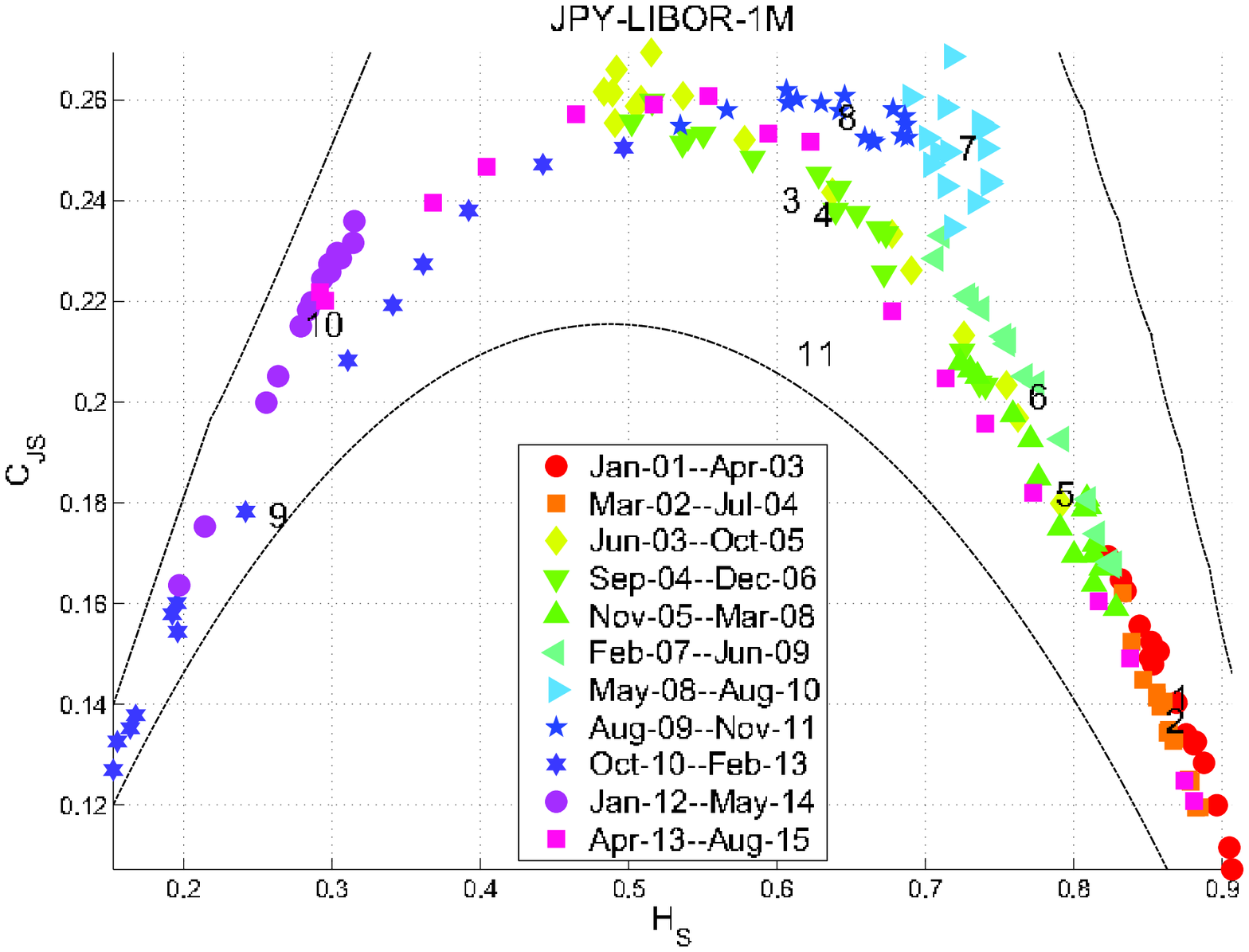}
        }
          \subfigure{%
           \label{fig:CECPjpy2M}
           \includegraphics[width=0.45\textwidth]{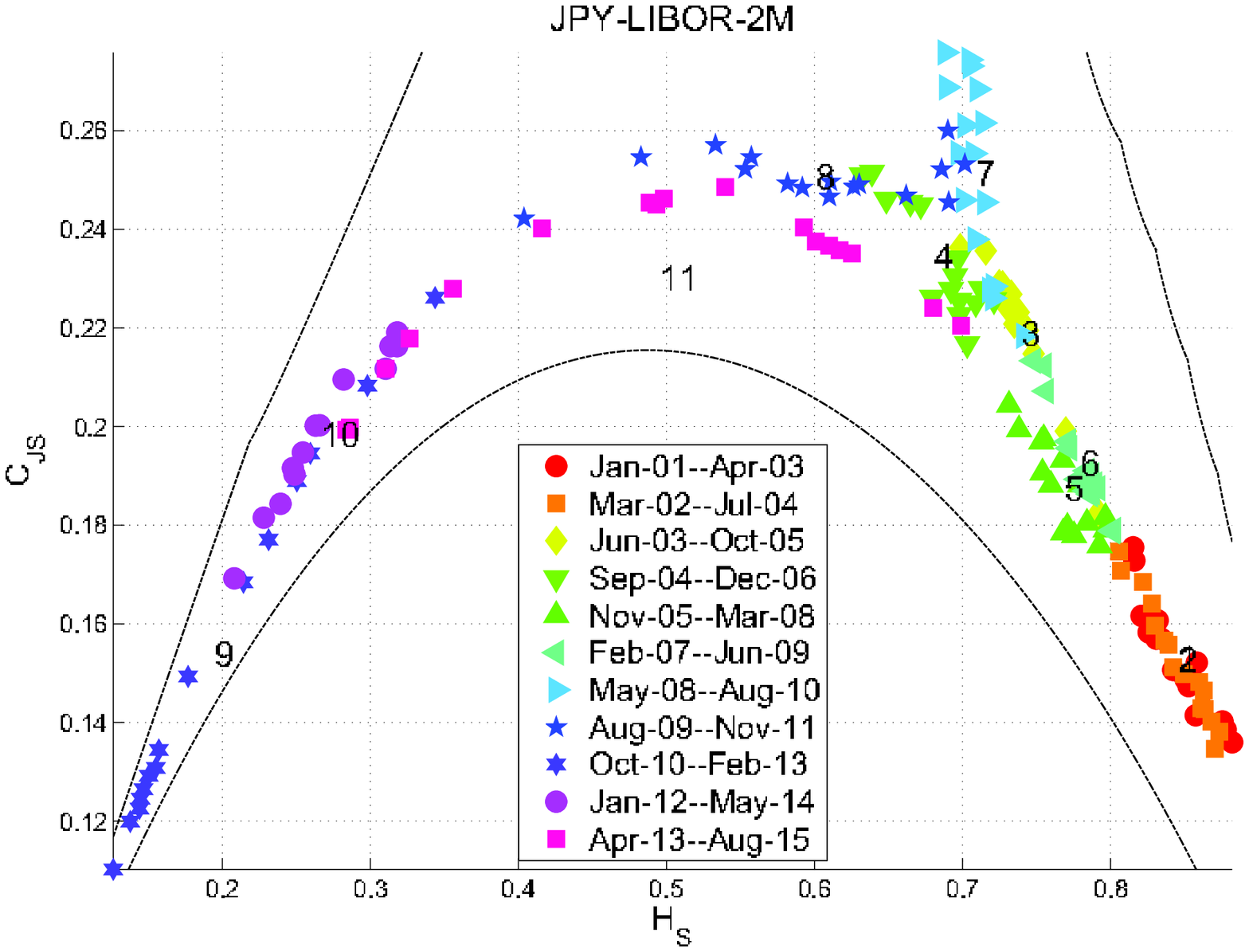}
        }%
}
    \caption{Complexity Entropy Causality Plane, with $D=4,\tau=1,\delta=20$ of JPY Libor for different maturities: overnight (O/N), one week (1W), one month (1M), two months (2M).
Numbers $\{1, \dots, 11\}$ are the central points of each of the clusters.
The solid lines represent the upper and lower bounds of the quantifiers as computed by Mart\'{\i}n {\it et al.\/} \cite{paper:martin2006}}
   \label{fig:CECP_JPYlibor1}
\end{figure}

\begin{figure}[!ht]
    \centering
{
%\\%------- End of the second row ----------------------%        
       \subfigure{%
          \label{fig:CECPjpy3M}
          \includegraphics[width=0.45\textwidth]{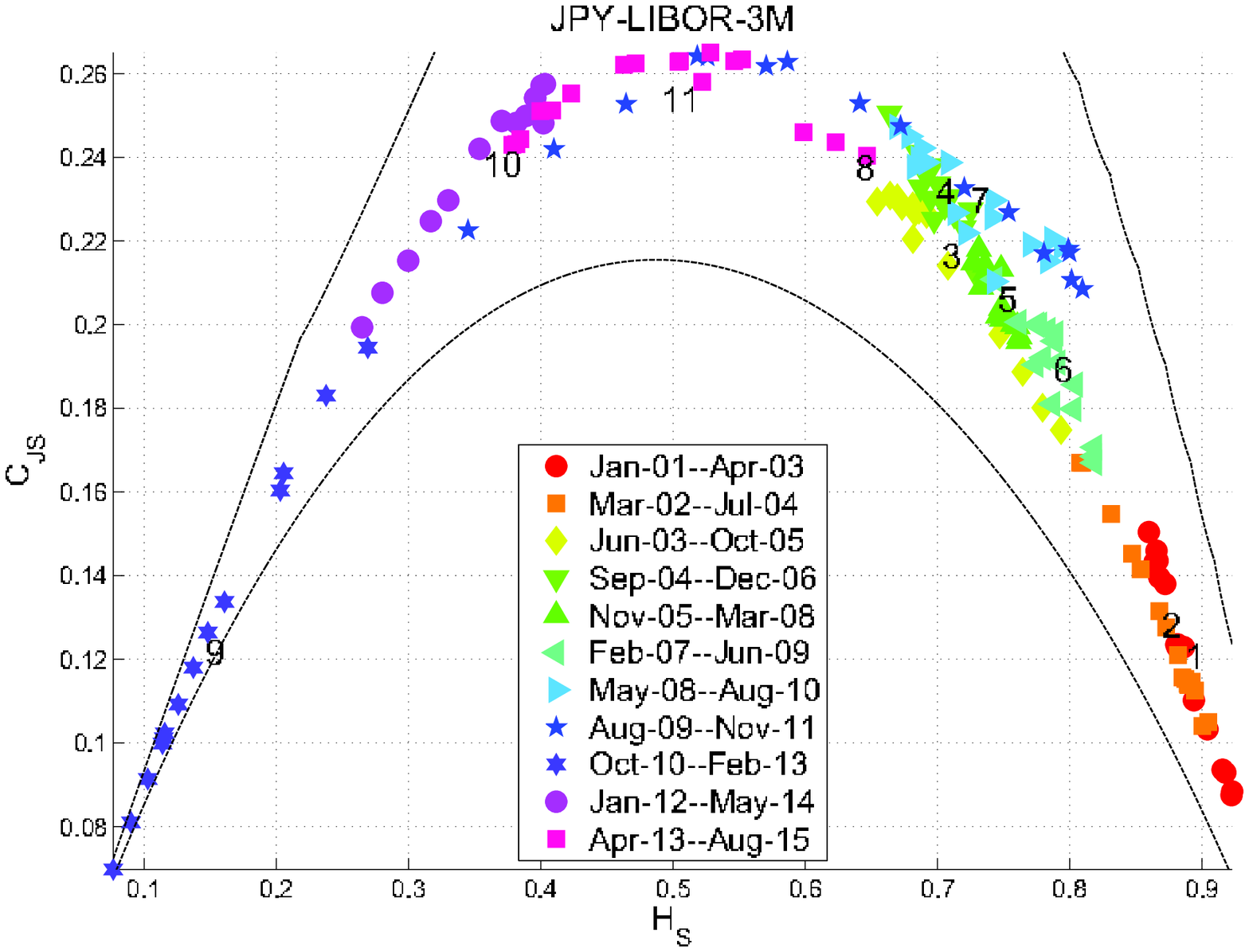}
       }
       \subfigure{%
          \label{fig:CECPjpy6M}
          \includegraphics[width=0.45\textwidth]{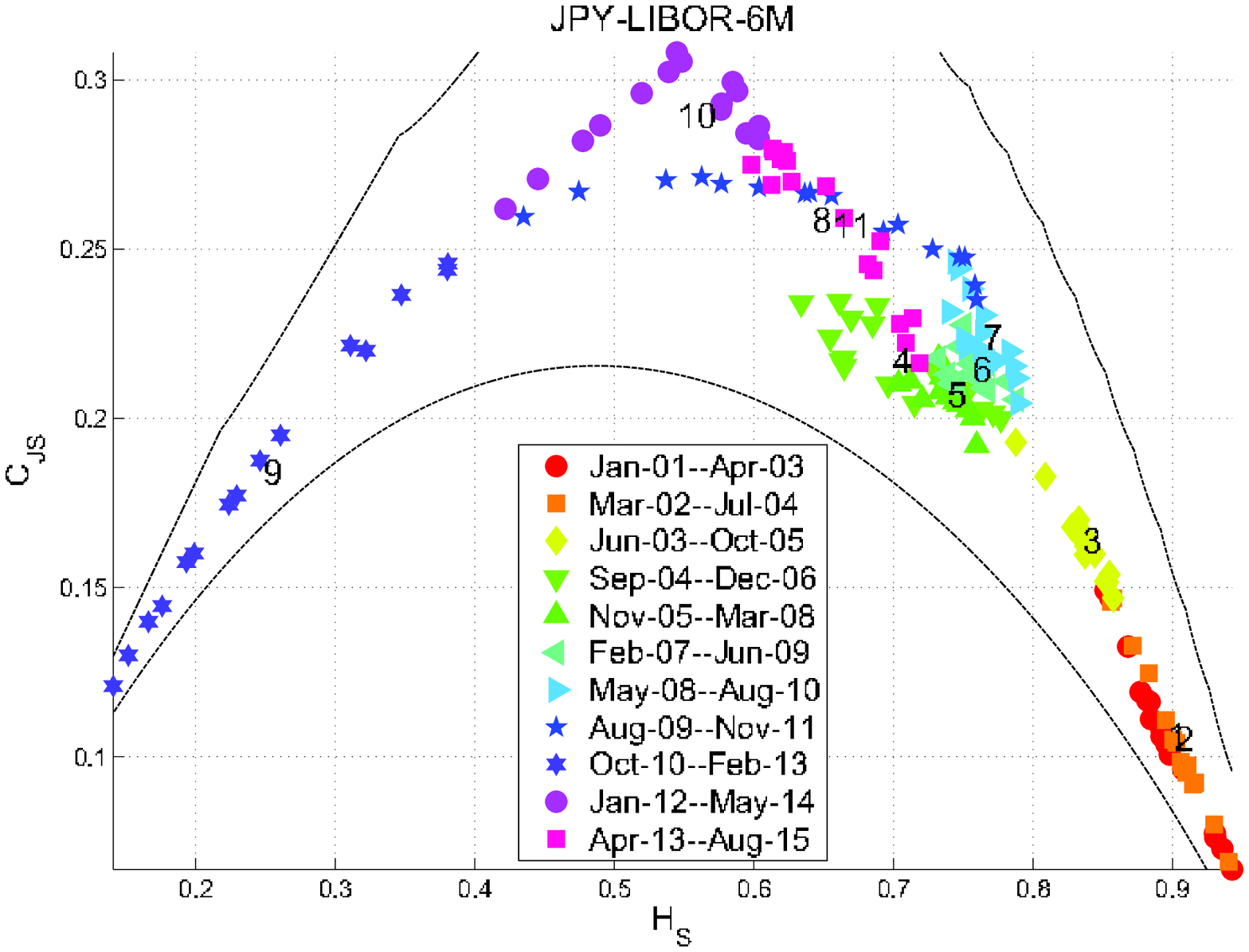}
      }
\\%------- End of the third row ----------------------%  
       \subfigure{%
          \label{fig:CECPjpy12M}
          \includegraphics[width=0.45\textwidth]{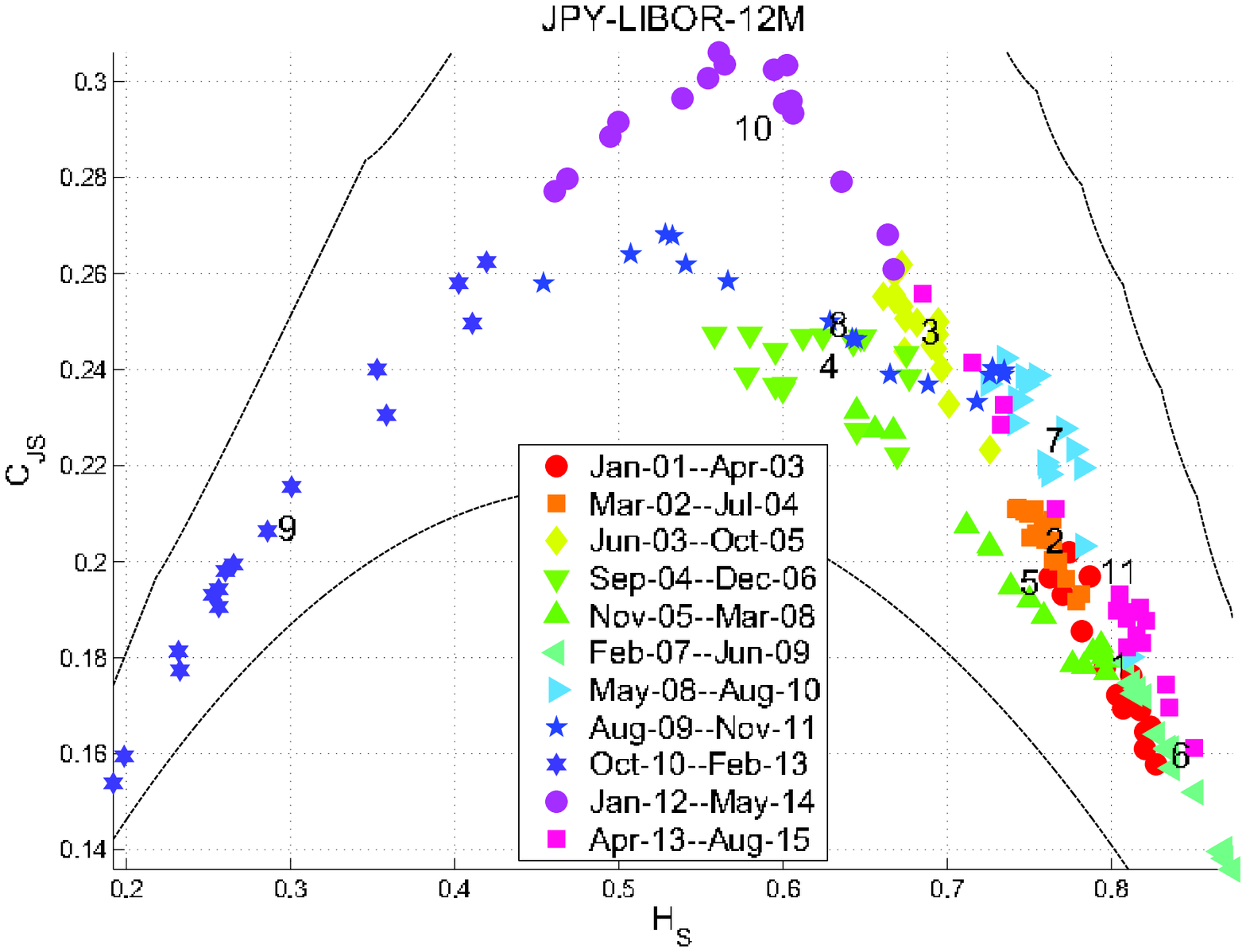}
       }
}
    \caption{Complexity Entropy Causality Plane, with $D=4,\tau=1,\delta=20$ of JPY Libor for different maturities (continuation): three months (3M), six months (6M) and twelve months (12M).
 Numbers $\{1 , \cdots , 11\}$ are the central points of each of the clusters.
The solid lines represent the upper and lower bounds of the quantifiers as computed by Mart\'{\i}n {\it et al.\/} \cite{paper:martin2006}}
   \label{fig:CECP_JPYlibor2}
\end{figure}

A simple observation of Figures \ref{fig:CECP_GBPlibor1} to \ref{fig:CECP_JPYlibor2} shows that we are facing a changing dynamic. The process governing interest rates does not seem to be stable over time. The reflection of this is that the position of the estimators changes radically in different temporal windows. However, this change is not random, but rather seems to follow a directed path. To make a more visual presentation, we have grouped the windows in 11 periods of 16 windows each (17 windows in the last period). So we can differentiate each period with a color and a different marker. Additionally, we have put a number to each period and we have located in the average values of entropy and complexity of that period. As a general rule, we can see that GBP, EUR and CHF Libor behaves very efficiently during the first three periods (years 2001-2005). Indeed, entropy is greater than 0.8 and less than 0.2 complexity. Period 4 appears to be a certain transition. Entropy decreases and complexity increases. This trend is deepened in subsequent periods, with periods 6, 7 and 8 being the most inefficient (years 2007-2012). Periods 9, 10 and 11  (years 2012-2015) show a return to the area of greatest informational efficiency.

A more detailed analysis by currency allows us to discover that not all maturities followed the same pattern. Indeed, the most affected are the maturities of 1, 2 and 3 months. At the other extreme, the least affected were maturities of overnight and 12 months. Further analysis should JPY Libor. The behavior is similar to other currencies, but all maturities have also been affected in the rate rigging.

Probably one of the reasons for the distinct behavior of JPY and the rest of the currencies is that Libor JPY is less used as a benchmark for pricing other financial instruments. On the other hand, the distinct behavior in the different maturities can be also explained by their use as a reference rate\footnote{see the use of the different Libor rate maturities and currencies as a reference rate for interest rate swaps and floating rate notes in Table C.2 in \cite{WheatleyReport}}.

We cannot discard that the financial crisis itself produced a disruption in the Libor market, making it less efficient. Its influence seems to depend on the nature of the financial assets under study. For example, \cite{BaGuMa12} report an asymmetric impact of the crisis in the long memory of corporate and sovereign bonds. However, it is at least a remarkable coincidence that the changes in informational efficiency is contemporary with the alleged manipulation, specially in some maturities. Additionally, the informational efficiency recovery begins when banks were fined by improper conduct. Moreover, our results agree with the finding in \cite{Barivieraetal2016}, that between 2007 and 2009 the Libor time series was more predictable than either before or after those years. 

\begin{figure}[!ht]
    \centering
{
        \subfigure{%
            \label{fig:Inef_GBP}
            \includegraphics[width=0.45\textwidth]{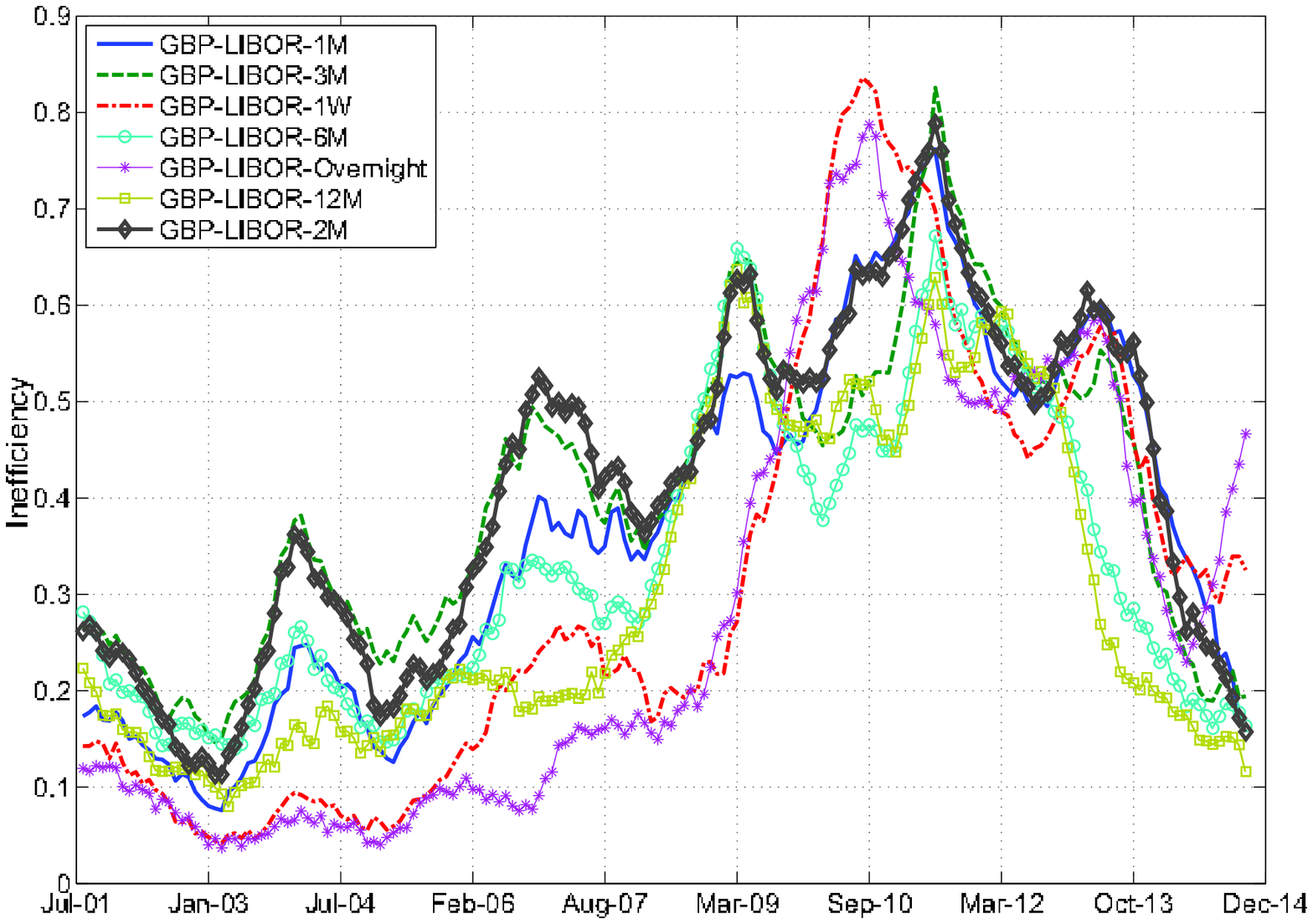}
        }%
        \subfigure{%
           \label{fig:Inef_EUR}
           \includegraphics[width=0.45\textwidth]{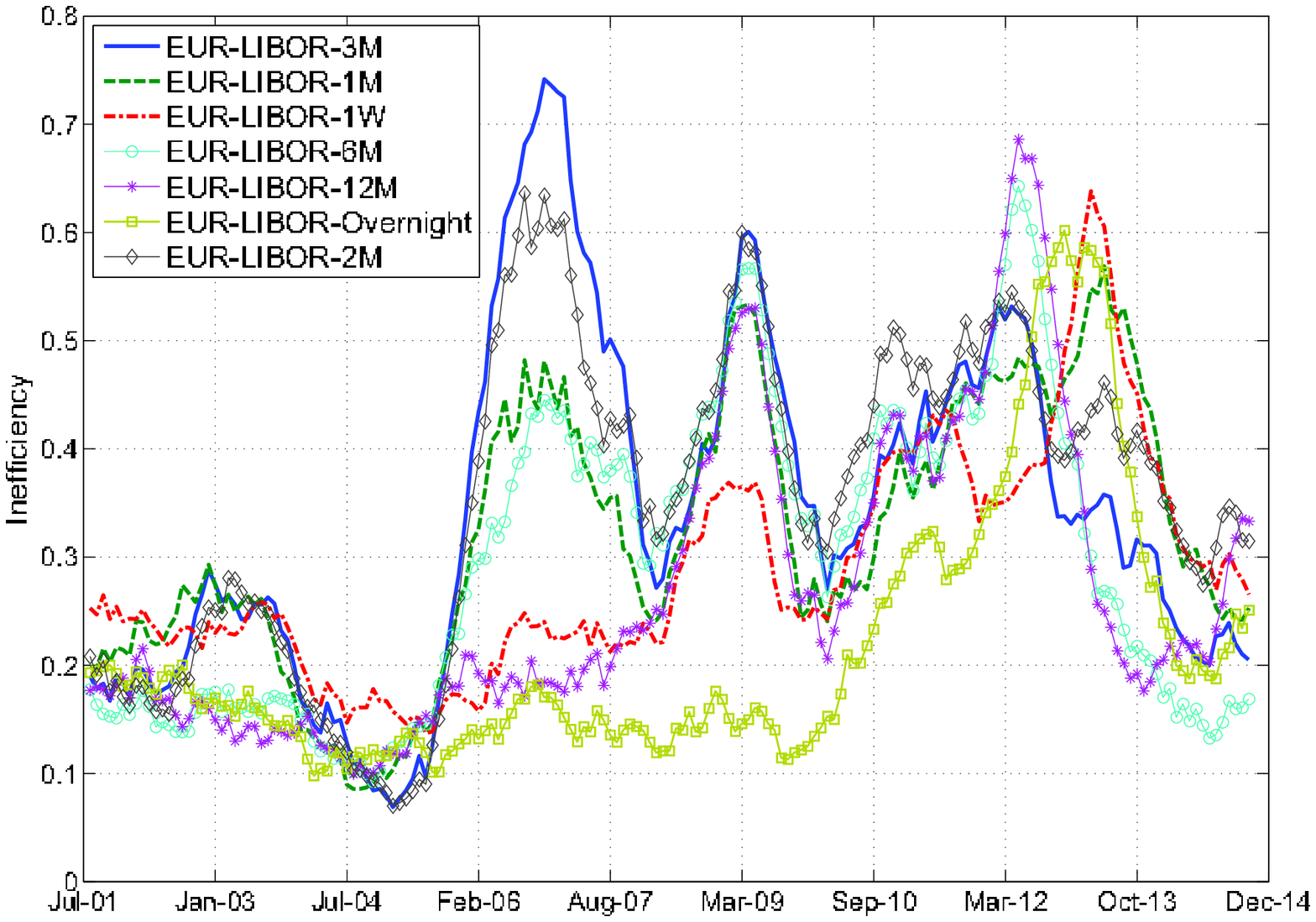}
        }
\\%  ------- End of the first row ----------------------%
        \subfigure{%
           \label{fig:Inef_CHF}
           \includegraphics[width=0.45\textwidth]{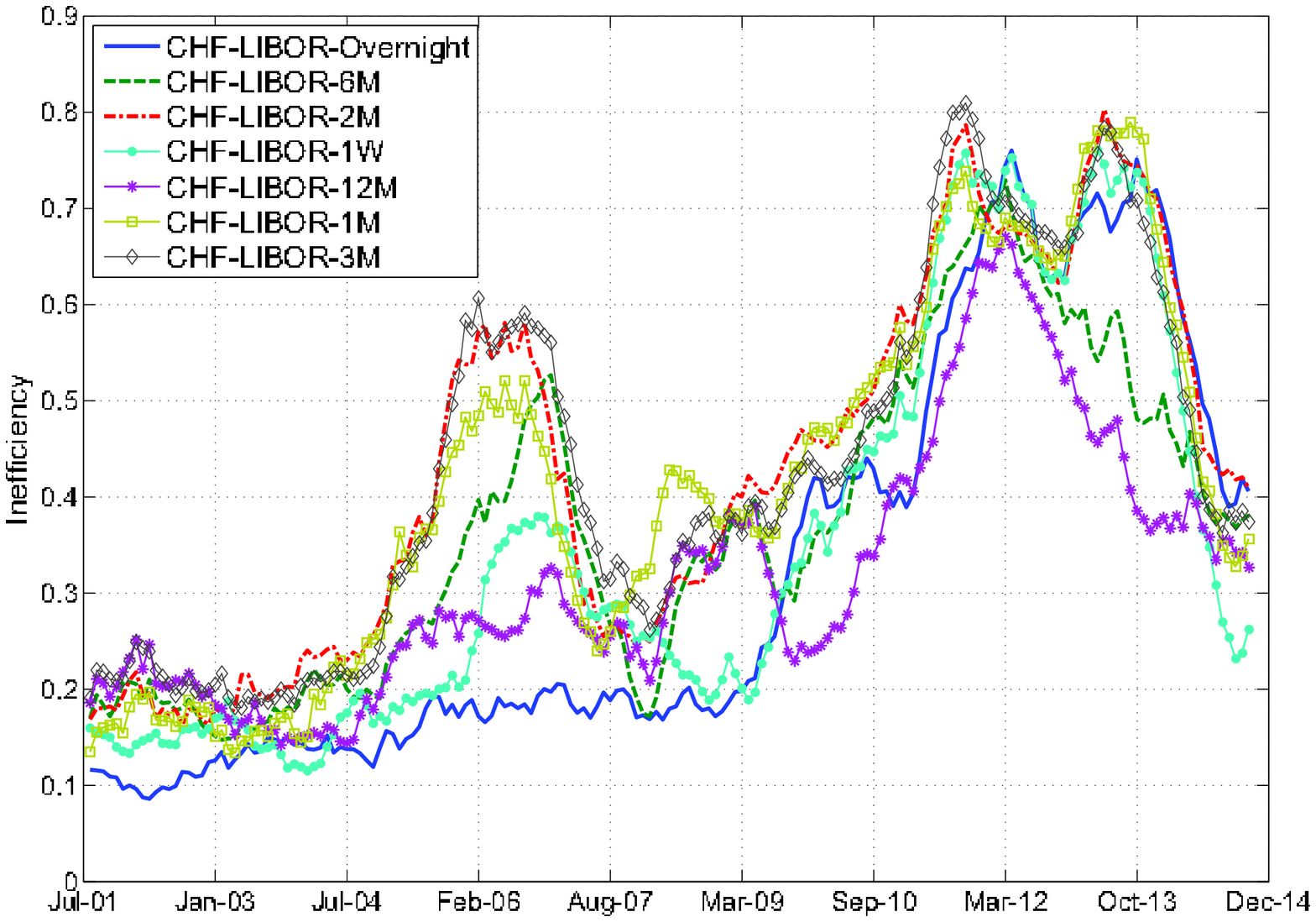}
        }
          \subfigure{%
           \label{fig:Inef_JPY}
           \includegraphics[width=0.45\textwidth]{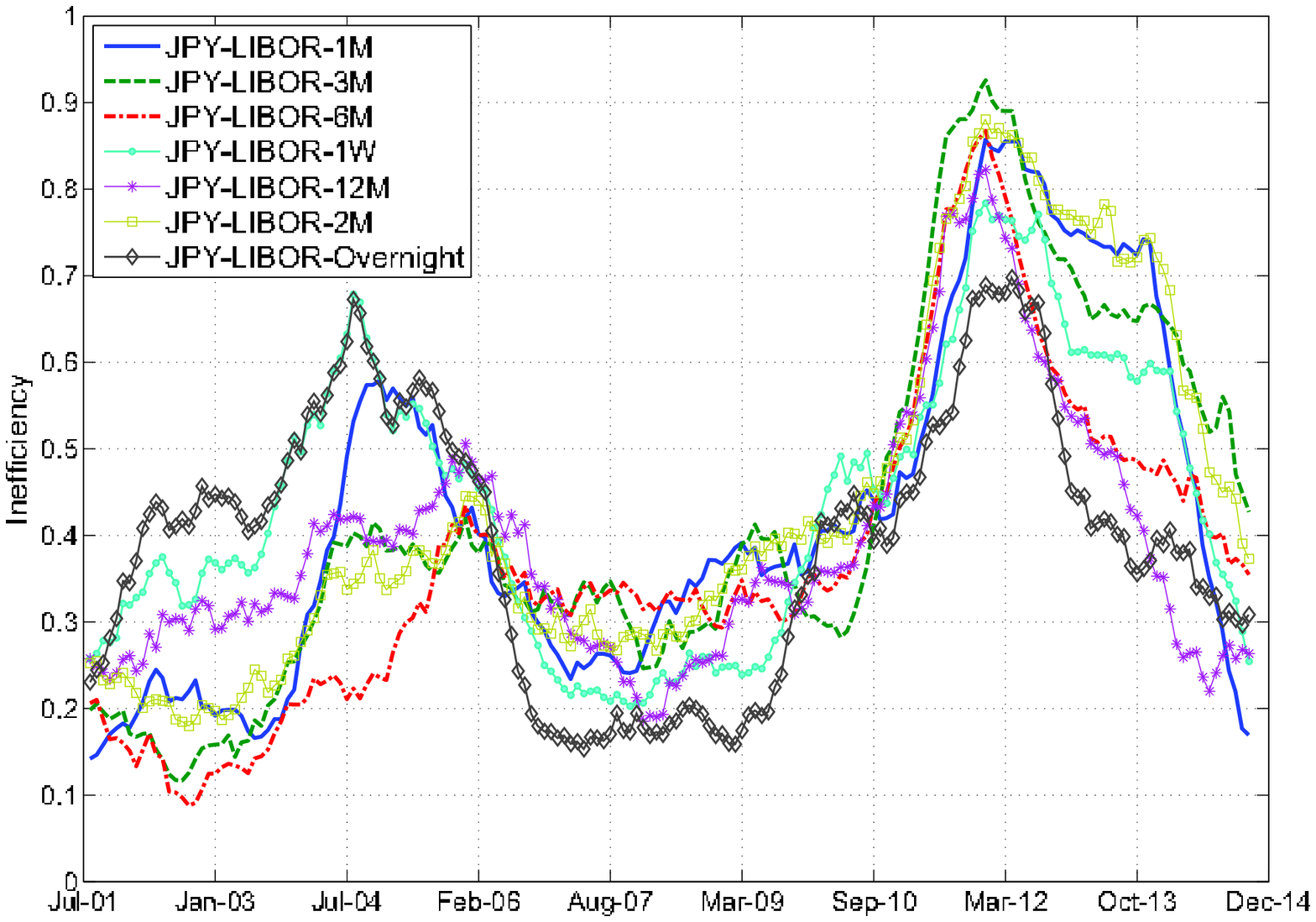}
        }%
}
    \caption{Inefficiency evolution for each currency and maturity of Libor rates, according to equation
 \ref{eq:inefficiency}}
   \label{fig:Inefficiency}
\end{figure}

In order to observe more clearly the temporal changes in informational efficiency, we compute the metric introduced in \cite{Bariviera2013epjb}:
\begin{equation}
\label{eq:inefficiency}
\text{Inefficiency}=+\sqrt{({\cal H}_S -1)^2+({\cal C}_{JS})^2}.
\end{equation}
This measure represents the Euclidean distance to the point ${\cal H}_S=1$ and ${\cal C}_{JS}=0$  , i.e. the maximal efficiency point.
The results can be observed in Figure \ref{fig:Inefficiency}.

%%%%%%% CECP PLANE OF GBP LIBOR  %%%%%%%%%%%%%%%%%%%

%%%%%%% CECP PLANE OF GBP LIBOR  %%%%%%%%%%%%%%%%%%%

%%%%%%% CECP PLANE OF EUR LIBOR  %%%%%%%%%%%%%%%%%%%

%%%%%%% CECP PLANE OF EUR LIBOR  %%%%%%%%%%%%%%%%%%%

%%%%%%% CECP PLANE OF CHF LIBOR  %%%%%%%%%%%%%%%%%%%

%%%%%%% CECP PLANE OF CHF LIBOR  %%%%%%%%%%%%%%%%%%%

%%%%%%% CECP PLANE OF JPY LIBOR  %%%%%%%%%%%%%%%%%%%

%%%%%%% CECP PLANE OF JPY LIBOR  %%%%%%%%%%%%%%%%%%%

%%%%%%% INEFFICIENCY EVOLUTION  %%%%%%%%%%%%%%%%%%%

\section{Conclusions \label{sec:Conclusions}}

This paper studies the 28 time series of Libor rates during the last 14 years. The information theory based symbolic analysis is known as Complexity-Entropy Causality Plane, a novel approach in financial economics. The use of the CECP allows the discrimination of different stochastic and chaotic regimes. We used  moving windows in order to introduce temporal dimension into our analysis. According to our results an abnormal movement of Libor time series arround the period of the 2007 financial crisis is detected. This alteration in the stochastic dynamics of Libor is contemprary of what press called ``Libor scandal'', i.e. the manipulation of interest rates carried out by several prime banks. We argue that our methodology is suitable as a market watch mechanism, as it makes visible the temporal redution in informational efficiency of the market.  Our results could be useful for regulatory authorities, since the procedure detailed in this paper could act as an early warning mechanism to detect unusual dynamics in the Libor market.

\section{Acknowledgements}
Lisana B. Martinez, M. Bel\'en Guercio and Osvaldo A. Rosso acknowledge financial support from Consejo Nacional 
de Investigaciones Cient\'{\i}ficas y T\'ecnicas (CONICET), Argentina.


\begin{thebibliography}{100}

 
%1  
\bibitem{MollenkampWhitehouse}
C. Mollenkamp, M. Whitehouse, 
Study casts doubt on key rate: WSJ analysis suggests banks may have reported flawed interest data for libor,
The Wall Street Journal, Thursday, 29 May, p.1, 2008.

%2 
\bibitem{Reuters2012}
Reuters, New york federal reserve knew about libor rate-fixing issues
as far back as 2007 and proposed changes but were ignored, 
The Daily Mail, July 10, 2012.

%3 
\bibitem{FTemail}
L. Saigol, 
Libor: The email trail, 
The Financial Times, http://www.ft.com/cms/s/0/cefd67a0-25df-11e3-aee8-
00144feab7de.html\#axzz3FNK2XAFe, 2013. URL: http://www.
ft.com/cms/s/0/cefd67a0-25df-11e3-aee8-00144feab7de.html\
\#axzz3FNK2XAFe.

%4 
\bibitem{HouSkeie}
D. Hou, D. R. Skeie, 
Libor: Origins, economics, crisis, scandal, and reform, 
Federal Reserve Bank of New York Staff Report No. 667, 2014.

%5 
\bibitem{TaylorWilliams2009}
J. B. Taylor, J. C. Williams, 
A black swan in the money market, 
American Economic Journal: Macroeconomics 1 (2009) 58-–83.

%6 
\bibitem{SniderYoule}
C. A. Snider, T. Youle, 
Does the libor reflect banks’ borrowing costs?,
Available at SSRN: http://ssrn.com/abstract=1569603, 2010. doi:10.2139/ssrn.1569603.

%7 
\bibitem{AbrantesMetz2011}
R. M. Abrantes-Metz, S. B. Villas-Boas, G. Judge, 
Tracking the libor rate, 
Applied Economics Letters 18 (2011) 893–-899.

%8 
\bibitem{Monticini20133}
A. Monticini, D. L. Thornton, 
The effect of underreporting on {LIBOR} rates, 
Journal of Macroeconomics 37 (2013) 345--348.

%9
\bibitem{BaiPerron1998}
J. Bai, P. Perron, 
Estimating and testing linear models with multiple structural changes, 
Econometrica 66 (1998) 47–-78.

%10 
\bibitem{EPJB2015}
A. F. Bariviera, M. Guercio, L. Martinez, O. Rosso,   
The (in)visible hand in the libor market: an information theory approach, 
The European Physical Journal B 88 (2015) 208.

%11
\bibitem{BarivieraRSTA2015} 
A. F. Bariviera, M. Guercio, L. Martinez, O. Rosso, 
A permutation information theory tour through different interest rate maturities: the libor case, 
Philosophical Transactions of the Royal Society of London A:
Mathematical, Physical and Engineering Sciences 373 (2015) 20150119.

%12 
\bibitem{WheatleyReport}
M. Wheatley, 
The Wheatley Review of LIBOR: Final Report, 
Independent report, H.M. Treasury, 2012.

%13 
\bibitem{TheilLeenders65}
H. Theil, C. T. Leenders, 
Tomorrow on the Amsterdam stock exchange,
The Journal of Business 38 (1965) 277-–284.

%14 
\bibitem{Fama65entropy}
E. F. Fama, 
Tomorrow on the New York stock exchange, 
The Journal of Business 38 (1965) 285-–299.

%15 
\bibitem{Dryden68}
M. M. Dryden, 
Short-term forecasting of share prices: an Information Theory approach, 
Scottish Journal of Political Economy 15 (1968) 227-–249.

%16 
\bibitem{PhilippatosNawrocki73}
G. C. Philippatos, D. N. Nawrocki, 
The behavior of stock market aggregates: Evidence of dependence on the american stock exchange, 
Journal of Business Research 1 (1973) 101-–114.

%17 
\bibitem{PhilippatosWilson74}
G. C. Philippatos, C. J. Wilson, 
Information theory and risk in capital markets, 
Omega 2 (1974) 523–-532.

%18 
\bibitem{Risso08}
W. A. Risso, 
The informational efficiency and the financial crashes, 
Research in International Business and Finance 22 (2008) 396–-408.

%19 
\bibitem{Risso09}
W. A. Risso, 
The informational efficiency: the emerging markets versus the developed markets, 
Applied Economics Letters 16 (2009) 485–-487.

%20 
\bibitem{OrtizCruz12}
A. Ortiz-Cruz, E. Rodriguez, C. Ibarra-Valdez, J. Alvarez-Ramirez, 
Efficiency of crude oil markets: Evidences from informational entropy analysis, 
Energy Policy 41 (2012) 365–-373.

%21 
\bibitem{book:shannon1949}
C. E. Shannon, W. Weaver, 
The Mathematical Theory of Communication, 
University of Illinois Press, Champaign, IL, 1949 

%22 
\bibitem{FeldmanCrutchfield98}
D. P. Feldman, J. P. Crutchfield, 
Measures of statistical complexity: Why?, 
Physics Letters A 238 (1998) 244–-252.

%23
\bibitem{LMC95} 
R. L\'opez-Ruiz, H. L. Mancini, X. Calbet, 
A statistical measure of complexity, 
Physics Letters A 209 (1995) 321–-326.

%24 
\bibitem{Martin2003}
M. Mart\'{\i}n, A. Plastino, O. Rosso, 
Statistical complexity and disequilibrium, 
Physics Letters A 311 (2003) 126--132.

%25 
\bibitem{Lamberti2004119}
P. W. Lamberti, M. T. Mart\'{\i}n, A. Plastino, O. A. Rosso, 
Intensive entropic non-triviality measure, 
Physica A 334 (2004) 119–-131.

%26
\bibitem{paper:martin2006}
M. T. Mart\'{\i}n, A. Plastino, O. A. Rosso, 
Generalized statistical complexity measures: Geometrical and analytical properties, 
Physica A 369 (2006) 439–-462.

%27
\bibitem{Soriano2011a} 
M. C. Soriano, L. Zunino, O. A. Rosso, I. Fischer, C. R. Mirasso, 
Time scales of a chaotic semiconductor laser with optical feedback under the lens of a permutation 
information analysis, 
IEEE J. Quantum Electron. 47 (2011) 252–-261.

%28
\bibitem{Zunino2010a}
L. Zunino, M. C. Soriano, I. Fischer, O. A. Rosso, C. R. Mirasso,
Permutation-information-theory approach to unveil delay dynamics from time-series analysis, 
Physical Review E 82 (2010) 046212.

%29 
\bibitem{ZuninoCausality10}
L. Zunino, M. Zanin, B. M. Tabak, D. G. P\'erez, O. A. Rosso,
Complexity-entropy causality plane: A useful approach to quantify the stock market inefficiency, 
Physica A: Statistical Mechanics and its Applications 389 (2010) 1891-–1901.

%30
\bibitem{ZuninoPermutation11}
L. Zunino, B. M. Tabak, F. Serinaldi, M. Zanin, D. G. P\'erez, O. A. Rosso, 
Commodity predictability analysis with a permutation information theory approach, 
Physica A: Statistical Mechanics and its Applications 390 (2011) 876-–890.

%31
\bibitem{RossoNoise07}
O. A. Rosso, H. A. Larrondo, M. T. Mart\'{\i}n, A. Plastino, M. A. Fuentes,
Distinguishing noise from chaos, 
Physical Review Letters 99 (2007) 154102.

%32
\bibitem{BandtPompe02}
C. Bandt, B. Pompe, 
Permutation entropy: A natural complexity measure for time series, 
Physical Review Letters 88 (2002) 174102.

%33 
\bibitem{Saco2010}
P. M. Saco, L. C. Carpi, A. Figliola, E. Serrano, 330 O. A. Rosso, 
Entropy analysis of the dynamics of el ni\~no southern oscillation during the holocene, 
Physica A: Statistical Mechanics and its Applications 389 (2010) 5022--5027.

%34
\bibitem{paper:keller2005}
K. Keller,M. Sinn, 
Ordinal analysis of time series, 
Physica A: Statistical Mechanics and its Applications 356 (2005) 114–-120.

%35 
\bibitem{Zunino2012}
L. Zunino, A. F. Bariviera, M. B. Guercio, L. B. Martinez, O. A. Rosso,
On the efficiency of sovereign bond markets, 
Physica A 391 (2012) 4342–-4349.

%36 
\bibitem{SerinaldiZuninoRosso2013}
F. Serinaldi, L. Zunino, O. A. Rosso, 
Complexity-entropy analysis of daily stream flow time series in the continental united states, 
Stochastic Environmental Research and Risk Assessment 28 (2014) 1685–-1708.

%37
\bibitem{Zanin2012} 
M. Zanin, L. Zunino, O. A. Rosso, D. Papo, 
Permutation entropy and its main biomedical and econophysics applications: A review, 
Entropy 14 (2012) 1553–-1577.

%38
\bibitem{Samuelson65} 
P. A. Samuelson, 
Proof that properly anticipated prices fluctuate randomly, 
Industrial Management Review 6 (1965) 41–-49.

%39
\bibitem{BaGuMa12}
A. F. Bariviera, M. B. Guercio, L. B. Martinez, 
A comparative analysis of the informational efficiency of the fixed income market in seven european countries, Economics Letters 116 (2012) 426–-428.

%40
\bibitem{Barivieraetal2016}
A. F. Bariviera, M. T. Mart\'{\i}in, A. Plastino, V. Vampa, 
LIBOR troubles: Anomalous movements detection based on maximum entropy, 
Physica A: Statistical Mechanics and its Applications 449 (2016)401--407.

%41
\bibitem{Bariviera2013epjb}
A. F. Bariviera, L. Zunino, M. B. Guercio, L. B. Martinez, O. A. Rosso, 
Revisiting the European sovereign bonds with a permutation information-theory approach, 
The European Physical Journal B 86 (2013) 509. 

\end{thebibliography}
\end{document}